\documentclass[12pt]{article}
\usepackage{epsf,amssymb,latexsym}
\usepackage{rotating}
\usepackage{natbib}
\usepackage{letterspace}
\usepackage{times}
\usepackage{mathptmx} 
\newcommand{\grtsim}{\raisebox{-.6ex}{$\stackrel{\textstyle{>}}{\sim}$}}
\renewcommand{\baselinestretch}{2.0} 

\begin{document}

\renewcommand{\baselinestretch}{1.0} 

\begin{center}
\Huge
Galileo In-Situ Dust Measurements in Jupiter's Gossamer Rings
\end{center}

\vspace*{5mm}
\begin{center}
\Large
Harald~Kr\"uger$^{1,2}$,
Douglas P. Hamilton$^{3}$,
Richard Moissl$^{1}$,
and Eberhard Gr\"un$^{2,4}$
\end{center}

\vspace*{7mm}
\begin{center}
\large
$^1$
Max-Planck-Institut f\"ur Sonnensystemforschung, \\
Max-Planck-Str. 2, \\
37191 Katlenburg-Lindau, \\
Germany \\
E-Mail: krueger@mps.mpg.de \\[6pt]
$^2$
Max-Planck-Institut f\"ur Kernphysik,\\
Postfach 103980, 69029 Heidelberg, Germany\\[6pt]
$^3$
Astronomy Department,\\
University of Maryland, \\ 
College Park, \\
MD\,20742-2421, \\
USA\\[6pt]
$^4$ Laboratory for Atmospheric and Space Physics, University of Colorado, \\ 
Boulder, CO, 80303-7814, USA \\[6pt]
\end{center}

\vspace*{5mm}
\begin{tabular}{lr}
Manuscript pages: & 62 \\
Figures:          & 13 \\
Tables:           & 4 \\
\end{tabular}

\vspace*{5mm}
\begin{center}
\Large
{\it Icarus},  submitted

\thispagestyle{empty}
\vfill
\today\\
\end{center}


\newpage
\vspace*{20mm}
\begin{center}
{\sl Proposed Running Head:}\\[1mm]
IN-SITU DUST MEASUREMENTS IN JUPITER'S GOSSAMER RINGS
\end{center}

\vspace*{15mm}
\begin{center}
{\sl Corresponding author:}\\[1mm]
{\sf Harald Kr\"uger}\\[-1mm]
Max-Planck-Institut f\"ur Sonnensystemforschung, \\[-1mm]
Max-Planck-Str. 2, \\[-1mm]
37191 Katlenburg-Lindau, Germany\\
E-mail: krueger@mps.mpg.de
\end{center}

\newpage
\renewcommand{\baselinestretch}{2.0} 
\newpage

\vspace*{20mm}

\medskip

\begin{abstract}
  Galileo was the first artificial satellite to orbit Jupiter. During
  its late orbital mission the spacecraft made two passages through
  the giant planet's gossamer ring system.  The impact-ionization dust
  detector on board successfully recorded dust impacts during both
  ring passages and provided the first in-situ measurements from a
  dusty planetary ring. During the first passage -- on 5 November 2002
  while Galileo was approaching Jupiter - dust measurements were
  collected until a spacecraft anomaly at $\rm 2.33\, R_J$ (Jupiter
  radii) just 16~min after a close flyby of Amalthea put the
  spacecraft into a safing mode.  The second ring passage on 21
  September 2003 provided ring dust measurements down to about $2.5\,
  \rm R_J$ and the Galileo spacecraft was destroyed shortly thereafter
  in a planned impact with Jupiter.
  In all, a few thousand dust impacts were counted with the instrument
  accumulators during both ring passages, but only a total of 110
  complete data sets of dust impacts were transmitted to Earth.
  Detected particle sizes range from about 0.2 to $\rm 5\, \mu m$,
  extending the known size distribution by an order of magnitude
  towards smaller particles than previously derived from optical
  imaging \citep{showalter2008}. The grain size distribution increases
  towards smaller particles and shows an excess of these tiny motes 
  in the Amalthea gossamer ring compared to the Thebe ring.
  The size distribution for the Amalthea ring
  derived from our in-situ measurements for the small grains agrees very 
  well with the one obtained from images for large grains. 
  Our analysis shows that particles
  contributing most to the optical cross-section are about $\rm 5\,
  \mu m$ in radius, in agreement with imaging results. The
  measurements indicate a large drop in particle flux immediately
  interior to Thebe's orbit and some detected particles seem to be on
  highly-tilted orbits with inclinations up to $20^{\circ}$.  
  Finally, the faint Thebe ring extension
  was detected out to at least $\rm 5\,R_J$, indicating that grains
  attain higher eccentricities than previously thought.  The drop
  interior to Thebe, the excess of submicron grains at Amalthea, 
  and the faint ring extension indicate that grain dynamics is
  strongly influenced by electromagnetic forces. These findings can
  all be explained by a shadow resonance as detailed by
  \cite{hamilton2008}.
\end{abstract}

{\bf Keywords:}\\
\hspace*{3cm}dust\\
\hspace*{3cm}Jupiter, satellites\\
\hspace*{3cm}Jupiter, rings\\
\hspace*{3cm}planetary rings


\newpage

\sloppy

\section{Previous Imaging Results}

All four giant planets of our Solar System are surrounded by huge
tenuous ring systems which contain mostly micrometer- and
submicrometer-sized dust particles \citep{burns2001}. In these rings,
dust densities are so low that particle collisions are negligible, and
grain dynamics is substantially perturbed by non-gravitational
forces. The 'dusty' rings are interesting and valuable counterpoints
to the collisionally dominated opaque and dense rings of Saturn and
Uranus which are populated primarily by macroscopic centimeter- to
meter-sized objects.

\begin{figure}
\epsfxsize=0.8\hsize 
\epsfbox{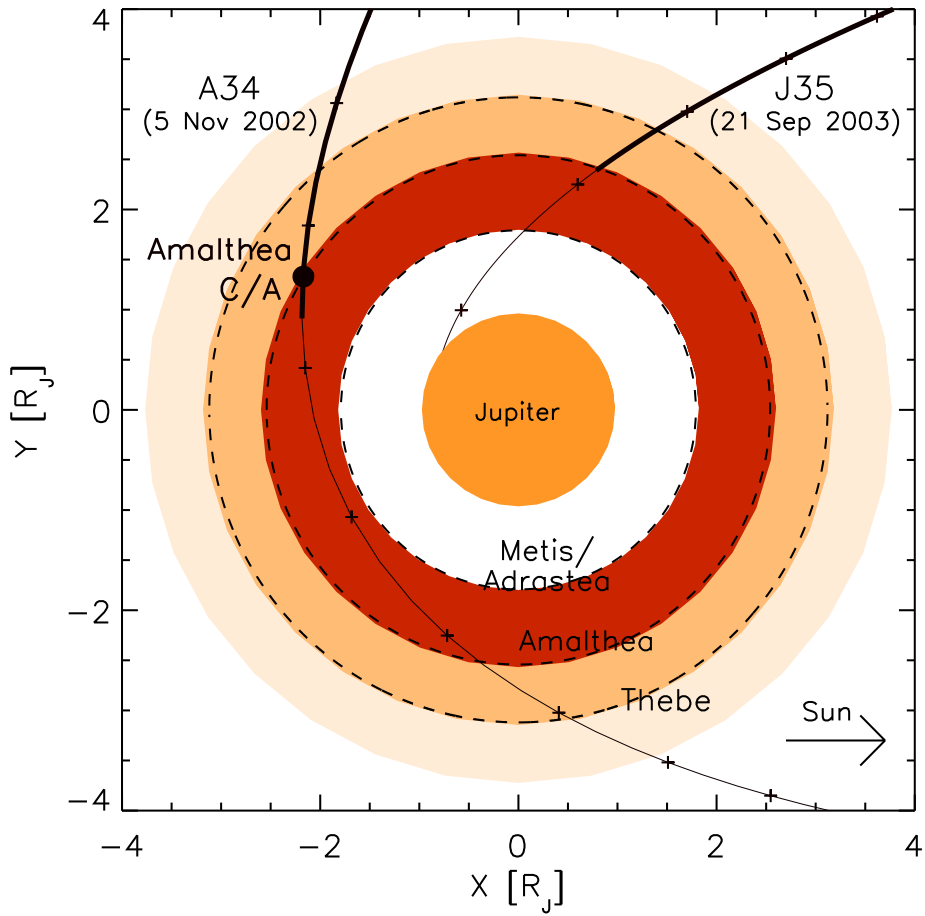}
         \caption{\label{orbitplot1}
Projection of Galileo's trajectory through the gossamer ring region on 
5 November 2002 (Galileo orbit A34)
and 21 September 2003 (orbit J35) onto Jupiter's equatorial plane. The 
orbits of the small moons Thebe, Amalthea,
Adrastea and Metis are indicated by dashed lines. Crosses indicate 1-hour 
time intervals. The different gossamer ring regions are
highlighted. Galileo's closest 
approach to Amalthea occurred on 5 November 2002 at 06:19 UTC (indicated by
a filled circle). Thick solid sections of Galileo's trajectory indicate time 
periods when dust data were obtained.  
}
\end{figure}

\begin{figure}
\parbox{\hsize}{
\hspace{17.5mm}         
\parbox{\hsize}{
\vspace{2cm}            
\epsfxsize=0.732\hsize
\epsfbox{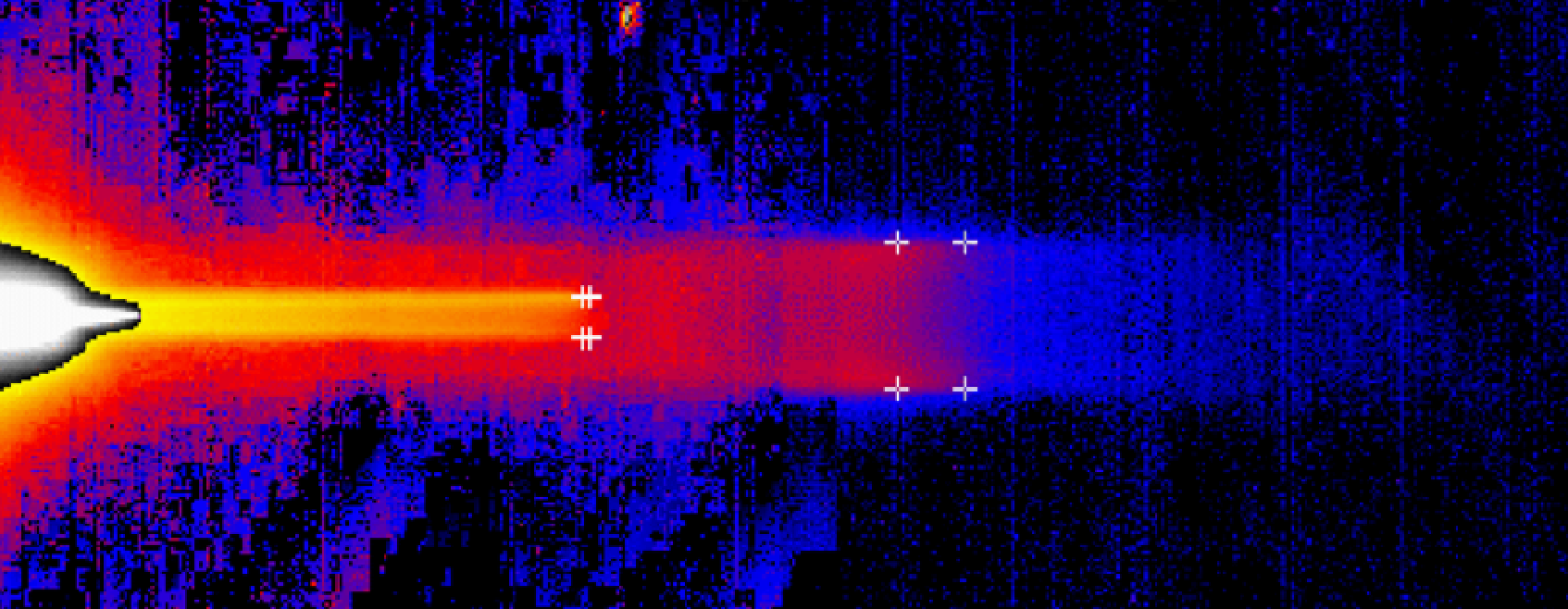}
\vspace{-45.7mm}          
}
\parbox{\hsize}{
\hspace{-1.5mm}           
\begin{turn}{90}
\epsfxsize=0.52\hsize
\epsfbox{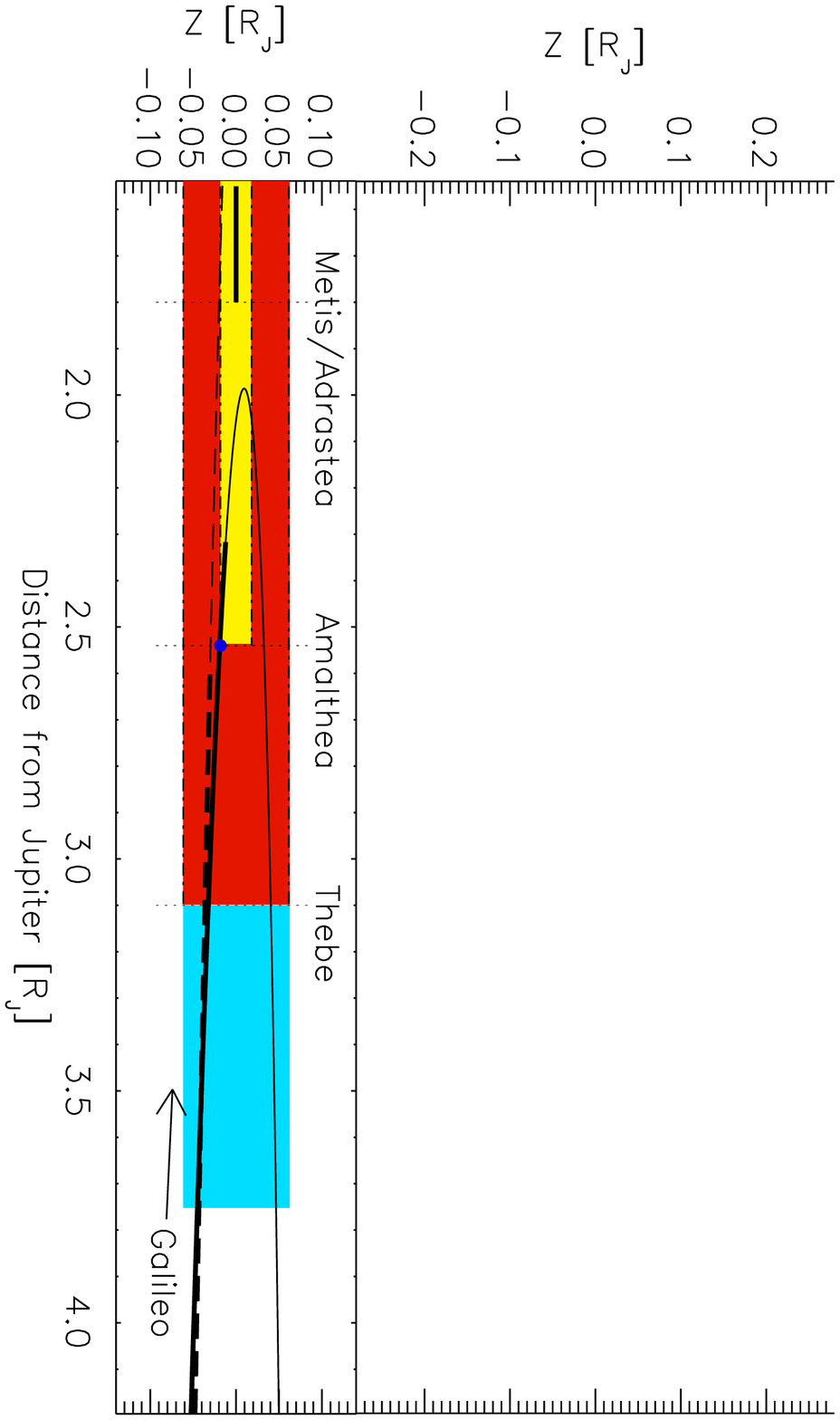}
\end{turn}
}
}
\caption{\label{orbitplot2} {\it Top:} Mosaic of Galileo images of
  Jupiter's gossamer rings taken when the spacecraft was very nearly
  in the ring plane \citep[from][]{burns1999}.  The halo and main ring
  are overexposed (solid white) at the left hand side of the image.
  To the right are the Amalthea ring (shown in light grey) and the
  Thebe ring is (shown in darker grey).  Crosses mark the four
  extremes of the radial and vertical motions of Amalthea and Thebe as
  caused by their eccentric and inclined orbits.  A very faint
  extension reaches out beyond Thebe's orbit. {\it Bottom:} Galileo's
  trajectories during the ring passages on 5 November 2002 (solid
  line) and 21 September 2003 (dashed line).  The sections where dust
  data were collected during both passages are highlighted as thick
  lines.  The approximate locations of the moons' orbits are indicated
  by vertical dashed lines and Amalthea's position during closest
  approach on 5 November 2002 is marked by a filled circle.  
  (from \citet{hamilton2008}).}
\end{figure}

Jupiter's ring system was investigated with remote imaging from the
Earth and from the Voyager, Galileo and Cassini spacecraft, revealing
significant structure in the ring: at least four components have been
identified \citep{ockert-bell1999, burns1999, depater1999}: the main
ring, interior halo and two gossamer rings.  The small moons Metis,
Adrastea, Amalthea and Thebe are embedded in the ring system and act as
sources of ring dust via meteoroid impact erosion of their
surfaces \citep{burns1999}.  The faint gossamer rings appear to extend primarily
inward from the orbit of Amalthea and Thebe (Figures~\ref{orbitplot1}
and \ref{orbitplot2}). In addition, the vertical limits of each moon's
slightly inclined orbit very closely match the vertical extensions of
these two rings \citep{ockert-bell1999}. These observations imply a
close relationship between the rings and embedded moonlets.  Outside
the orbit of Thebe, a swath of faint material is seen out to about $
\rm 3.75\, R_J$ (Jupiter radius, $\rm R_J = 71,492~km$) distance from
the planet.  Beyond this distance, the rings fade slowly into the
background. 
Normal optical depths are about $ 10^{-6}$ for the main ring and halo,
and about 100 -- 1000 times less for the Amalthea ring and Thebe rings.
Analysis of the few gossamer ring images implies particle radii of
$\rm 5 - 10\,\mu m$ with additional contributions from larger material
\citep{showalter2008}. In this paper, we show that smaller grains are
also present in large numbers.

\marginpar{\fbox{Figs.~\ref{orbitplot1} and ~\ref{orbitplot2}}}

The simplest picture of particle dynamics in the ring implies that
dust grains ejected from the surfaces of each moon would rapidly
disperse in longitude and nodal angles while maintaining their initial
inclinations \citep{burns1999}.  As such material evolves inward under
Poynting-Robertson drag, it would naturally produce the two
overlapping rings with rectangular profiles. Support for this
interpretation comes from the fact that both gossamer rings show
concentrations at the vertical extremes, where particles on inclined
orbits spend most of their time.  The extension of Thebe's gossamer
ring beyond Thebe's orbit, however, violates this simple and elegant
picture and has been attributed to an electromagnetic process
involving Jupiter's intense magnetic field by \citet{hamilton2008}.

\section{Galileo In-Situ Dust Measurements}


The Galileo spacecraft was the first artificial satellite of Jupiter,
circling the giant planet between 1996 and 2003. Near the end of the
mission, the spacecraft passed directly through the rings twice, on 5
November 2002 and 21 September 2003, offering a unique opportunity for
in-situ studies of planetary rings. The in-situ dust detector on board
\citep{gruen1992a} counted several thousand dust impacts during both
ring passages, and the full data sets, consisting of impact direction,
charge amplitudes, rise times, etc., for 110 separate impacts were
transmitted to Earth.  The first ring passage included a close flyby
at Amalthea with a closest approach distance of 244~km, just outside
the Hill sphere of this jovian moon.  The flyby provided an improved
mass estimate for the satellite, with an implied density of $\sim
0.8$g\,cm$^{-3}$ \citep{anderson2005}.

Galileo's traversal of Jupiter's gossamer rings provided the first
in-situ measurements of a dusty planetary ring. In-situ dust
measurements nicely complement imaging, providing important additional
information about the physical properties of the dust environment.  In
particular, in-situ measurements constrain dust spatial densities
along the spacecraft trajectory as well as grain masses, size
distributions, impact speeds and grain dynamics.

In this paper we present and analyse the complete in-situ dust
measurements obtained during both Galileo gossamer ring passages.  We
analyse grain impact directions and impact rates and derive dust
number densities and grain size distributions from the measurements.
We interpret results in terms of the gossamer rings' structure and the
dynamics of charged ring particles.

\subsection{Dust Detection Geometry}

\label{sec_geometry}

\begin{figure}
\epsfxsize=0.7\hsize
\epsfbox{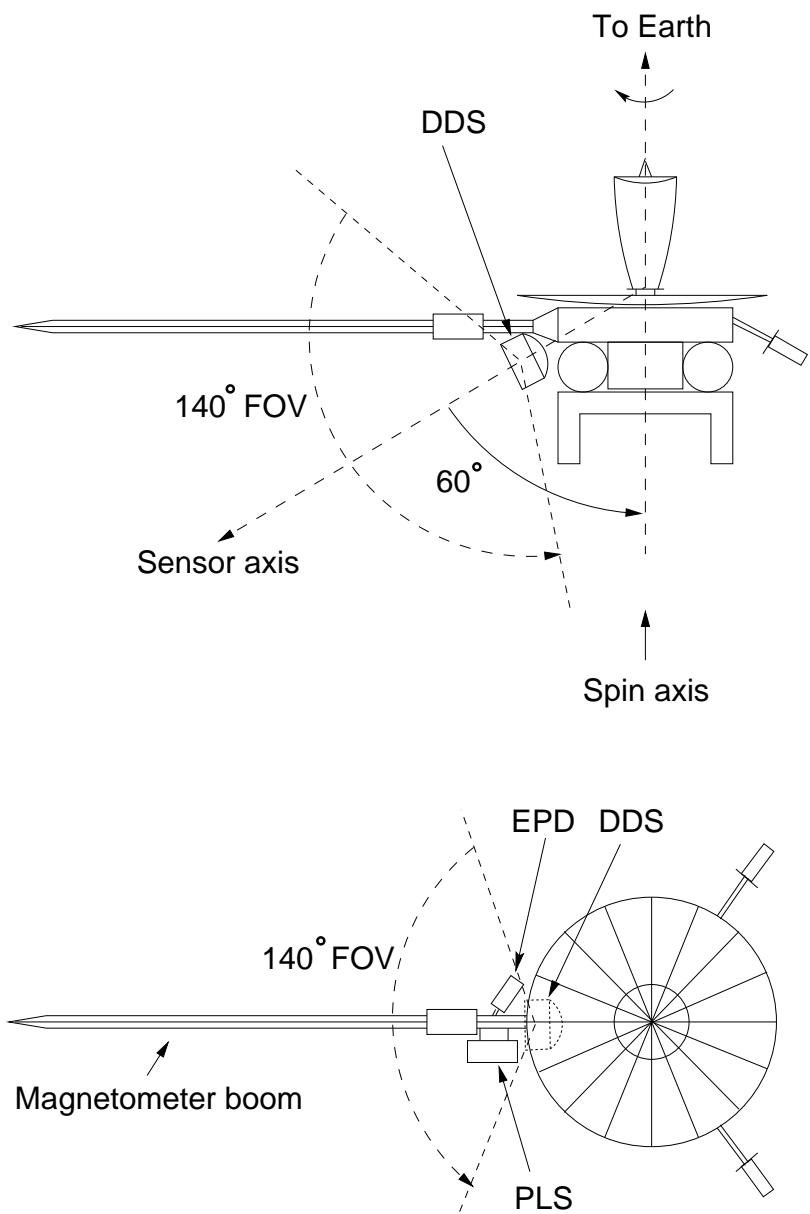}
\caption[Galileo spacecraft configuration (schematic)] {Galileo
  spacecraft configuration (schematic).  {\it Top:} Side view; {\it
    Bottom:} Top view.  The dust detector (DDS) is mounted directly
  underneath the magnetometer (MAG) boom \citep{kivelson1992}. The
  sensor field-of-view (FOV) is shown by dashed lines. The locations
  of the Plasma Instrument (PLS) \citep[PLS;][]{frank1992} and the
  Energetic Particles Detector (EPD) \citep[EPD;][]{williams1992},
  which partially obscure the DDS FOV, are also indicated.  }
\label{fig_galileo}
\end{figure}

Galileo was a dual spinning spacecraft with an antenna that pointed
antiparallel to the positive spin axis. The antenna 
usually pointed towards Earth. 
The Dust Detector System (DDS) was mounted on the spinning section of
Galileo underneath the magnetometer boom \citep{kivelson1992}, with
the sensor axis offset by $60^{\circ}$ from the positive spin axis.
Figure~\ref{fig_galileo} shows a schematic view of the Galileo
spacecraft and the geometry of dust detection.

\marginpar{\fbox{Fig.~\ref{fig_galileo}}}

The rotation angle, $\Theta$, measured the viewing direction of the
dust sensor at the time of a dust impact.  During one spin revolution
of the spacecraft, $\Theta$ scanned through a complete circle of
$360^{\circ}$. At $\Theta \simeq 90^{\circ}$ and $\simeq 270^{\circ}$
the sensor axis lay nearly in the ecliptic plane, and at $0^{\circ}$
it was close to the ecliptic north direction.  Rotation angles are
taken positive around the negative spin axis of the spacecraft which 
points towards Earth. This
is done to easily compare Galileo spin angle data with those taken by
Ulysses, which, unlike Galileo, has its positive spin axis pointed
towards Earth \citep{gruen1995a}.

The field-of-view (FOV) of the dust sensor target was $140^{\circ}$.
Due to the offset of $60^{\circ}$ between the sensor axis and the
spacecraft spin axis, over one spacecraft spin revolution, the sensor
axis scanned the surface of a cone with $120^{\circ}$ opening angle
centered on the anti-Earth direction.  Dust particles that arrived
from within $10^{\circ}$ of the positive spin axis (anti-Earth
direction) could be detected at all rotation angles $\Theta$, whereas
those that arrived with angles between $10^{\circ}$ and $130^{\circ}$
from the positive spin axis could be detected over only a limited
range of rotation angles. In the frame fixed to the spacecraft, we
define the {\em impact angle} between the impact velocity and the
sensor axis as $\phi$, and the angle between the impact velocity and the
spacecraft's anti-Earth spin axis as $\psi$.

Figure~\ref{fig_galileo} shows that the magnetometer boom
\citep[MAG;][]{kivelson1992} was in the field of view of the dust
sensor. The Energetic Particles Detector \citep[EPD;][]{williams1992}
and the Plasma Instrument \citep[PLS;][]{frank1992} partially obscured
the FOV of the dust sensor as well (Figure~\ref{ddsfov}). In other
words, at certain spacecraft rotation angles $\Theta$, particles
approaching at angles with respect to the spacecraft spin axis $\psi
\geq 90^{\circ}$ hit the boom and these Galileo instruments instead of
the sensor target.
The effect of this obscuration was first recognized in measurements 
of the jovian dust stream particles \citep{krueger1999c}. 

\marginpar{\fbox{Fig.~\ref{ddsfov}}}

\subsection{Dust Impact and Noise Identification}

\label{sec_identification}

Dust grains hitting the sensor target generated a plasma cloud of 
evaporating grain and target material.
For each impact, three independent measurements of the resulting plasma
cloud were used to derive the impact speed $ v$ and the mass $ m$ of
the particle: the electron signal, an ion signal, and a channeltron
signal \citep{gruen1992a}.  The charge $ Q$ released upon impact onto
the target is roughly described by the relation \citep{goeller1989}
\begin{equation} Q \varpropto m \cdot v\,^{3.5} .
\label{equ_charge}
\end{equation}
The dust instrument was empirically calibrated in the speed range 
2 to $\rm 70\,km\,s^{-1}$. Furthermore, the coincidence times of the three charge
signals together with the charges themselves are used to sort each
impact into one of four classes. Class~3 impacts have three
charge signals, two are required for class~2 and class~1 events,
and only one for class~0 \citep{baguhl1993b,gruen1995a,krueger1999a}.
In addition to the four classes, the dust data were categorised into
six amplitude ranges of the impact-generated ion charge, each range 
covering one order of magnitude in charge 
\citep[here denoted by AR1 to AR6;][]{gruen1995a}. Hence, taking
the classes and amplitude ranges together,
the dust data were grouped into $4\times 6 = 24$ 
categories.

Class 3 signals, our highest quality, are real dust impacts while
class 0 events are mostly noise. Class~1 and class~2 events were true
dust impacts in interplanetary space \citep{baguhl1993a,krueger1999a}.
However, during Galileo's entire Jupiter mission from 1996 to 2002 --
while the spacecraft was in the inner jovian magnetosphere --
energetic particles from the jovian plasma environment caused enhanced
noise rates in class~2 and the lower quality classes.  By analysing
the properties of the Io stream particles and comparing them with the
noise events, the noise could be eliminated from the class~2 data
\citep{krueger1999c,krueger2005a}. In particular, most class~0 and
class~1 events detected in the jovian environment are probably noise.

Before the two ring flybys that are the subject of this paper, Galileo
had only once been within $\rm 6\,R_J$ of the planet, on approach in
December 1995. Due to uncertainty about the effects of Jupiter's harsh
radiation environment, the dust instrument was switched to a less
sensitive mode to protect it \citep{gruen1996c}. Accordingly, a very
low noise rate was measured. The instrument's sensitivity was later
increased, and for the duration of the mission, it recorded an
increasing noise level with decreasing distance to the planet.

We have tested the applicability of the noise identification scheme,
described in detail by \citet{krueger1999c,krueger2005a}, to the
near-Jupiter region and improved upon it. A modified noise
identification scheme was derived for the gossamer ring data
\citep{moissl2005}, showing that class~1 also contains likely
candidates for real dust impacts.  For class~2, AR1 only the
target-ion grid coincidence was used as a criterion for noise events
(EIC\,=\,0) while for the higher amplitude ranges (AR2-6) the scheme
of \citet{krueger2005a} was applied unchanged (i.e. [$\mathrm{EA - IA}
\leq 1$ or $\mathrm{EA - IA} \geq 7$] and $\mathrm{CA} \leq 2$; EA, IA
and CA are the digital values of the charge amplitudes measured on the
target, ion grid and channeltron, respectively -- see
\citet{gruen1995a} for a description of these parameters). For class~1
the following criterion for noise events was used independent of the
amplitude range of the event: [$\mathrm{EA - IA} \leq 2$ or
$\mathrm{EA - IA} \geq 9$] and $\mathrm{CA} \leq 2$. More details of
the noise identification in the gossamer ring data are described by
Kr\"uger et al. (in prep.).

We use this scheme throughout this paper to separate noise events from
true dust impacts.  Note that this noise removal technique uses
statistical arguments and is applicable to large data sets only;
individual dust impacts may be erroneously classified as noise and
vice versa.

\subsection{Instrument Operation and Data Transmission}

Galileo had a very low data transmission capability because of the
failure of its high-gain antenna to open completely. For the dust
measurements this meant that the full set of parameters measured
during a dust particle impact or noise spike could only be transmitted
to Earth for a limited number of events. The data sets of all other
events (whether noise or true impacts) were lost. All events (dust
{\em and} noise), however, were always counted with one of the 24
accumulators \citep{gruen1995a} as described in
Section~\ref{sec_identification}. This allows us to correct the dust
measurements for incomplete data transmission and to derive reliable
event rates.  In particular, no indications for unrecognized
accumulator overflows were seen in the data from both gossamer ring
passages as has been problematic for some other stages of the mission.

Galileo dust data could be read out from the instrument memory with
different rates \citep[see][for a description]{krueger2001a}. In order
to maximise the data transmitted from the two gossamer ring passages,
the read-out cycle was set to the fastest useful mode during the
respective passage.  For the ring passage on 5 November 2002 this
meant that dust data were read-out from the instrument memory and
written to the Galileo tape recorder in so-called record mode which
started at 02:44 UTC, i.e. 18\,min before Galileo crossed Io's orbit
during approach to Jupiter.  The latest data set measured in each
amplitude range was read-out at approximately one-minute intervals and
written to the onboard tape recorder for later transmission to Earth.
Hence, for impact rates up to $\sim 1\,\mathrm{min^{-1}}$ in each
amplitude range, all data sets could be transmitted to Earth. For
higher rates, a fraction of these data sets were lost.  This mode gave
the highest time resolution of the dust measurements at any time
during the mission; about 1~minute.  The completeness of the
transmitted data sets
varied between 100\,\% in the highest amplitude ranges (AR2-4) in the
faint ring extension beyond Thebe's orbit down to only 4\,\% for the
lowest amplitude range (AR1) in the more populated Amalthea ring.

Dust data were obtained in record mode during Galileo's approach to
Jupiter until a spacecraft anomaly (safing) on 5 November 2002 at
06:35 UTC prevented the collection of further data. This anomaly
occurred at a distance of $\rm 2.33\,R_J$ from Jupiter, 16\,min after
closest approach to Amalthea (at $\rm 2.54\,R_J$) and limited the
total period of dust measurements obtained from the gossamer rings to
about 100\,min.  Although the instrument continued to measure dust
impacts after the spacecraft anomaly, the data were not written to the
tape and, hence, most of them were lost. Only the data sets of a few
impact events which occurred in the ring region traversed by Galileo
after the spacecraft anomaly were obtained from a full memory readout
on 18 November 2002. These data, however, have only a low time
resolution of about 4.3~hours which is on the order of the duration of
the entire gossamer ring passage. Only the total number of events
(dust plus noise) in each amplitude range can be derived from the
accumulators for the ring region traversed after the spacecraft
anomaly.

During Galileo's second gossamer ring passage on 21 September 2003,
the dust data had to be transmitted to Earth immediately because the
spacecraft struck Jupiter and was destroyed less than an hour later.
Therefore, the dust instrument memory was read-out in the fastest mode
that allowed data to be transmitted in real time \citep[realtime
science mode; see][]{krueger2001a}. Unfortunately, time resolution in
this mode was only 7~minutes. The completeness of the transmitted data
was about 10\% in the faint Thebe ring extension and about 5\% in the
Thebe ring. The last data set from the Galileo dust instrument
received on Earth was read out from the dust instrument memory at
17:59\,UTC when the spacecraft was at a jovicentric distance of about
$\rm 2.5\,R_J$. Thus, data from this ring passage provided in-situ
dust measurements from the gossamer rings for a total period of about
60\,min with no measurements coming from within Amalthea's orbit.

The motion of Galileo through the gossamer rings together with the
readout frequency of the dust instrument memory defined the maximum
spatial resolution achievable with the ring measurements. During the
first ring passage, with 1~min readout frequency in record mode,
Galileo moved $\rm \sim 1,800\,km$ through the ring along its
trajectory between two adjacent instrument readouts. This corresponds
to a motion in radial distance of about 1,100~km (or $\rm
0.015\,R_J$). For the second ring passage the spatial resolution was
only about 14,000\,km or $\rm 0.2\,R_J$ (radial). The ring and the
Galileo trajectory are sketched in Figures~\ref{orbitplot1} and
\ref{orbitplot2} and the characteristics of both ring passages are
summarized in Table~\ref{tab_dustmeas}.

\marginpar{\fbox{Tab.~\ref{tab_dustmeas}}}

During the entire first ring passage a total of several thousand dust 
impacts were counted. Approximately 330 of these happened 
before the spacecraft safing at $\rm 2.33\,R_J$ inbound to Jupiter.
With the optimised noise identification scheme described in 
Section~\ref{sec_identification}
complete data sets of 90 true dust impacts were identified in the
Galileo recorded data from the region between $\rm 3.75\,R_J$ and $\rm
2.33\,R_J$.
During the second ring passage approximately 260 dust impacts were
counted down to $\rm 2.5\,R_J$ inbound to Jupiter. At this distance
dust data transmission ceased before Galileo hit Jupiter. 
20 data sets of dust impacts detected between
$\rm 3.75\,R_J$ and $\rm 2.5\,R_J$ were transmitted
to Earth.

\subsection{Mass and Speed Calibration}

\label{sec_aging}

Grain impact speeds and masses were usually derived from 
Equation~\ref{equ_charge} and an empirical calibration 
obtained in the laboratory \citep{gruen1995a}.
Analysis of the dust data measured during Galileo's entire Jupiter 
mission, however, revealed strong degradation of the instrument 
electronics which affected the speed and mass calibration. The
degradation was most likely caused by the harsh radiation environment 
in the inner jovian magnetosphere, and a detailed analysis was 
published by \citet{krueger2005a}. Here we recall only the most
significant results which are relevant for the 
gossamer ring measurements:
i) the sensitivity of the instrument for dust impacts and
noise dropped with time,
ii) the amplification of the charge amplifiers 
degraded, leading to reduced measured impact charge values,
iii) drifts in the charge rise times measured at the target and the ion
    collector lead
to prolonged rise time measurements,
iv) degradation of the channeltron required five increases of the channeltron
high voltage during the Galileo Jupiter mission,
v) no impact or noise event was registered in the highest ion charge amplitude 
ranges AR5 and AR6 after July 1999.
In particular,
ii) and iii) affect the mass and speed calibration of the dust
instrument. For dust measurements taken after the year 2000, masses
and speeds derived from the instrument calibration must be taken with
caution because the electronics degradation was severe. Only in cases
where impact speeds are known from other arguments, such as exist here
in the gossamer rings, can reliable
particle masses be derived.
This will be discussed in more detail in Section~\ref{sec_masses}.


\section{Results}

\subsection{Dust Impact Rates}                 \label{sec_rate}

In Figure~\ref{rate} we show examples of the impact rates 
measured during either gossamer ring passage of Galileo as derived from
the accumulators of the dust instrument. We show the rates for the
classes and amplitude ranges for which a sufficiently large number of
events were counted so that meaningful rate curves could be derived.

\marginpar{\fbox{Fig.~\ref{rate}}}

The rates measured in all categories (i.e. classes and ion amplitude
ranges) increased during approach to Jupiter. From the outer edge of
the Thebe ring extension until the time when the dust measurements
stopped in the Amalthea ring due to the spacecraft anomaly, the
increase was about two orders of magnitude in the lowest channels,
AR1, whereas it was only one order of magnitude in the higher channels
(AR2-4). This indicates a higher fraction of small particles in the
Amalthea ring than in the Thebe ring and the faint Thebe ring
extension.  In all channels, the highest rates occurred inside
Amalthea's orbit when the spacecraft crossed into the more densely
populated Amalthea ring.
No impacts were measured in the largest categories AR5 and AR6 during
both gossamer ring passages.

The instrument accumulators do not contain any information of whether
the counted events were due to noise or real dust impacts. Since several of 
the instrument channels were sensitive to noise ({\em cf.}
Section~\ref{sec_identification}) an empirical noise correction factor
had to be applied. This factor can only be derived from the data sets 
transmitted with their full information and it is taken as the ratio 
between the number of noise events and the total number of events
transmitted within a given time interval
\citep[dust plus noise; see also][]{krueger2001a}. Here, the noise rate
was calculated as the average over a 1\,hour interval.
The criteria for the identification of individual noise events in the
gossamer ring data are given in Section~\ref{sec_identification}.

The rate data from the first ring passage show a dip between Thebe's
and Amalthea's orbits. It is most obvious in the lowest amplitude
range AR1 where we have the highest number of counted events. The
event rate dropped by about a factor of two to five at this location,
and the measurements obtained for other particle sizes and during the
second ring passage are consistent with the existence of this dip. It
should be noted, however, that the noise rate in classes~1 and 2
exceeded 80\% during some periods of the ring passage so that the
noise removal lead to large uncertainties in the impact rate. Only
class~3, our highest quality class, was noise-free but unfortunately
the event rate detected in this category was normally too low to
construct a useful impact rate profile.  The data in the lowest
amplitude range alone are not convincing, however, the higher
channels, which are mostly noise-free, show a similar drop inside
Thebe's orbit. This is evident in class~2, AR4 from the first passage
(top right panel in Figure~\ref{rate}).  During the second ring
passage, a sufficiently large number of class~3, AR4 events were
transmitted so that an impact rate profile from this noise-free
channel could be constructed (bottom right panel in
Figure~\ref{rate}). This data also indicates a dip inside Thebe's
orbit.  Additional support for this interpretation comes from 
increased energetic particle fluxes measured in the dip 
region with the EPD
instrument onboard Galileo (Norbert Krupp, priv. comm.).
We therefore conclude that the dip in the impact rate is
real, implying a true drop in the dust number density in the
Thebe ring.  The consequences for grain dynamics and the ring
structure will be discussed in Section~\ref{sec_dynamics}.

An additional feature is the extension of the outer gossamer ring far
beyond its previously known outer edge at $\rm 3.75\,R_J$.
Interestingly, the impact rate profile for the smallest particles is
relatively flat beyond $\rm 3.75\,R_J$ whereas inside this distance it
increases towards Jupiter. These small submicron particles
do not scatter light well and so cannot be seen in optical images;
they may be in the process of escaping the gossamer rings as
predicted by \citet{hamilton1993a}.


During its first ring passage on 5 November 2002 Galileo had a close flyby 
of 100-km Amalthea at a closest approach distance of 244\,km from the moon's center.
Because the Amalthea gossamer ring is believed to be maintained by 
collisional ejecta from Amalthea itself, an increased dust impact rate
is to be expected in the close vicinity of this moon. 
Galileo detected ejecta dust clouds within the Hill spheres
of all four Galilean moons, but outside the Hill spheres there was no 
noticeable enhancement \citep{krueger1999d,krueger2003b}. 
Taking the recently determined mass of Amalthea \citep{anderson2005},
its Hill radius is $r_{\rm Hill} = 130\rm \,km$, only slightly larger
than the moon itself. Thus a spike in the dust flux was not expected,
and is not apparent in the $\sim 40$-second period that Galileo was within
500\,km of Amalthea.
Determining the role of Amalthea as both a source and sink for
gossamer ring dust grains requires detailed physical models of i) the
interplanetary impactor population and ii) ring particle
dynamics. This primarily theoretical task is beyond the scope of the current
paper.

\subsection{Grain Impact Direction}

\label{sec_rot}

Images of the gossamer rings taken with Galileo and Earth-based
telescopes imply that the orbits of the ring particles have very low
inclinations with respect to Jupiter's equatorial plane below
$1.5^{\circ}$, and that the majority of the grains move on
low-eccentric or even circular orbits
\citep{depater1999,ockert-bell1999,burns1999}. In order to calculate
the impact direction of the measured ring particles onto the sensor
target and the corresponding effective sensor area for these grains,
we assumed that the particles orbit Jupiter on circular prograde
trajectories with effectively zero inclination 

The only additional parameters necessary are the
spacecraft trajectory (state vectors) and spacecraft orientation. The
spacecraft trajectory is shown in Figures~\ref{orbitplot1} and
~\ref{orbitplot2}, and the spacecraft orientation is constrained by
the fact that the antenna pointed within $3^{\circ}$ of the Earth
direction during both passages of Galileo through the gossamer rings.

With these assumptions, for particles assumed to be on prograde circular 
orbits, we calculated the dust impact direction and the
corresponding sensor area. 
During the first ring passage, the angle
with respect to the spin axis $\psi$ varied by only $4^{\circ}$ in
the time interval of interest here when we obtained high-rate recorded
data from the ring region. 
In this interval the target area, averaged
over one spacecraft spin revolution, was $50 - 55\,\mathrm{cm^2}$.
During the second ring passage $\psi$ varied by about $10^{\circ}$ and
the sensor target area changed between 200 and $230\,\mathrm{cm^2}$.
For both passages the expected rotation
angle for particles orbiting Jupiter on prograde circular trajectories
was $\Theta \approx 90^{\circ}$, and that for retrograde trajectories
$\Theta \approx 270^{\circ}$.

The range of the rotation angle distribution $\Delta \Theta$
is determined by the sensor FOV which is nominally $140^{\circ}$. A
smaller FOV was found for a subset of the ten-nanometer-sized jovian
dust stream particle impacts \citep{krueger1999c}; we believe that
this reduction is due to the small sizes and rapid speeds of stream
particles. In the gossamer rings, by contrast, we expect a larger than
nominal effective FOV; recent analysis of Galileo and Ulysses dust
data showed that the sensor FOV for particles much larger than the
jovian dust streams population is almost $180^{\circ}$ because the
inner sensor side wall showed a sensitivity for dust impacts
comparable to that of the target itself
\citep{altobelli2004a,willis2004,willis2005}. We therefore consider an
extended FOV for the analysis of gossamer ring particles.

\begin{figure}
\epsfxsize=0.5\hsize
\epsfbox{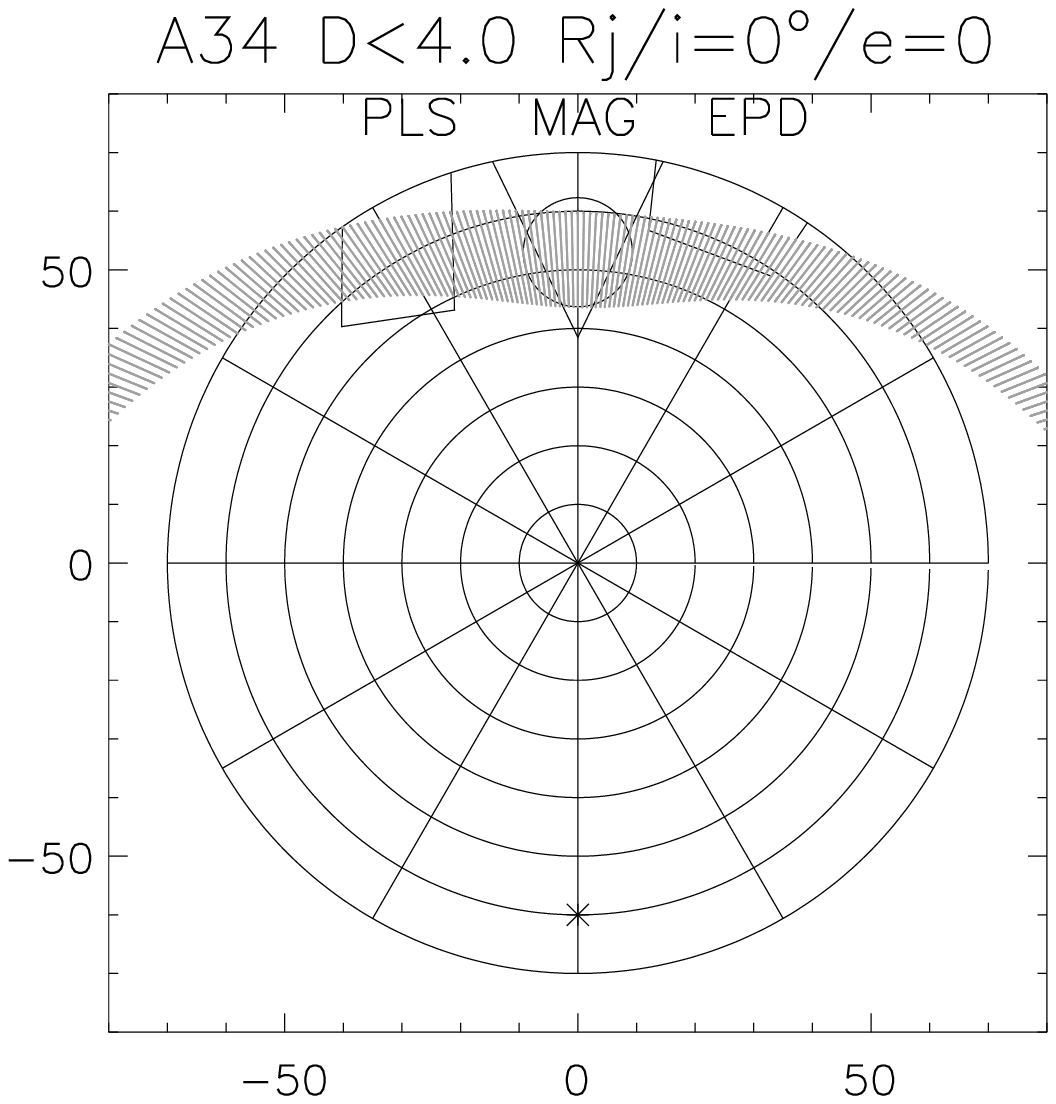}
\epsfxsize=0.5\hsize
\epsfbox{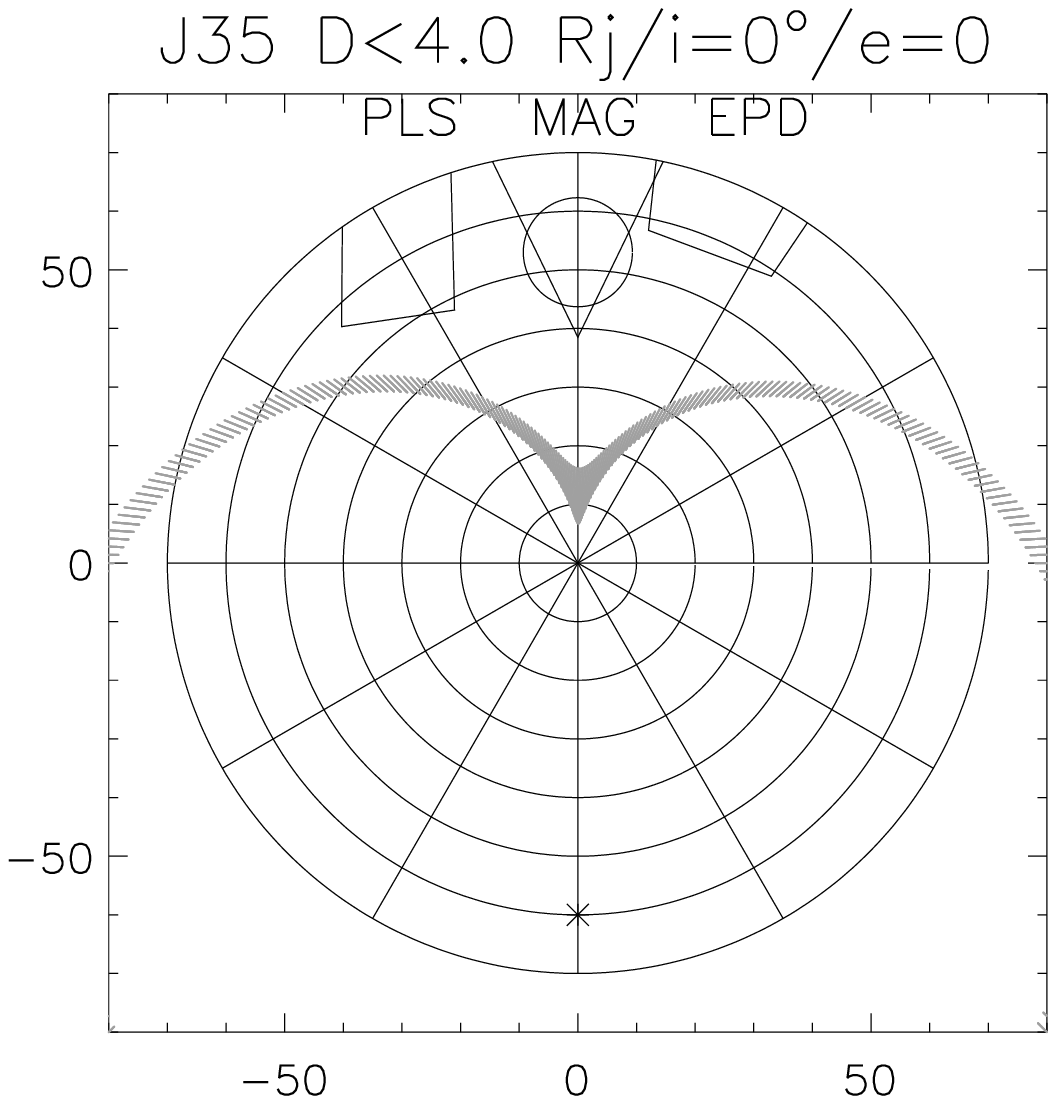}
\caption{\label{ddsfov} Dust instrument FOV and obscuration by the
  magnetometer boom, the PLS and the EPD instruments for an imaginary
  observer looking outward from the center of the sensor target. {\em Left:}
  first ring passage on 5 November 2002; {\em right:} second passage
  on 21 September 2003. Concentric circles denote the angular distance $\phi$
  from the sensor axis in $10^{\circ}$ steps. The spacecraft spin axis
  is at $\rm \phi = 60^{\circ}$ towards the bottom (marked by an
  asterisk). The shaded areas show the modelled range scanned by ring
  particles on circular prograde orbits during each ring passage
  \citep{moissl2005}. The width of the shaded areas is due to the
  variation of the angle $\psi$ between the impact velocity and the anti-Earth
  spin-axis during the motion of Galileo through the ring.  }
\end{figure}

\begin{figure}
\epsfxsize=0.5\hsize
\epsfbox{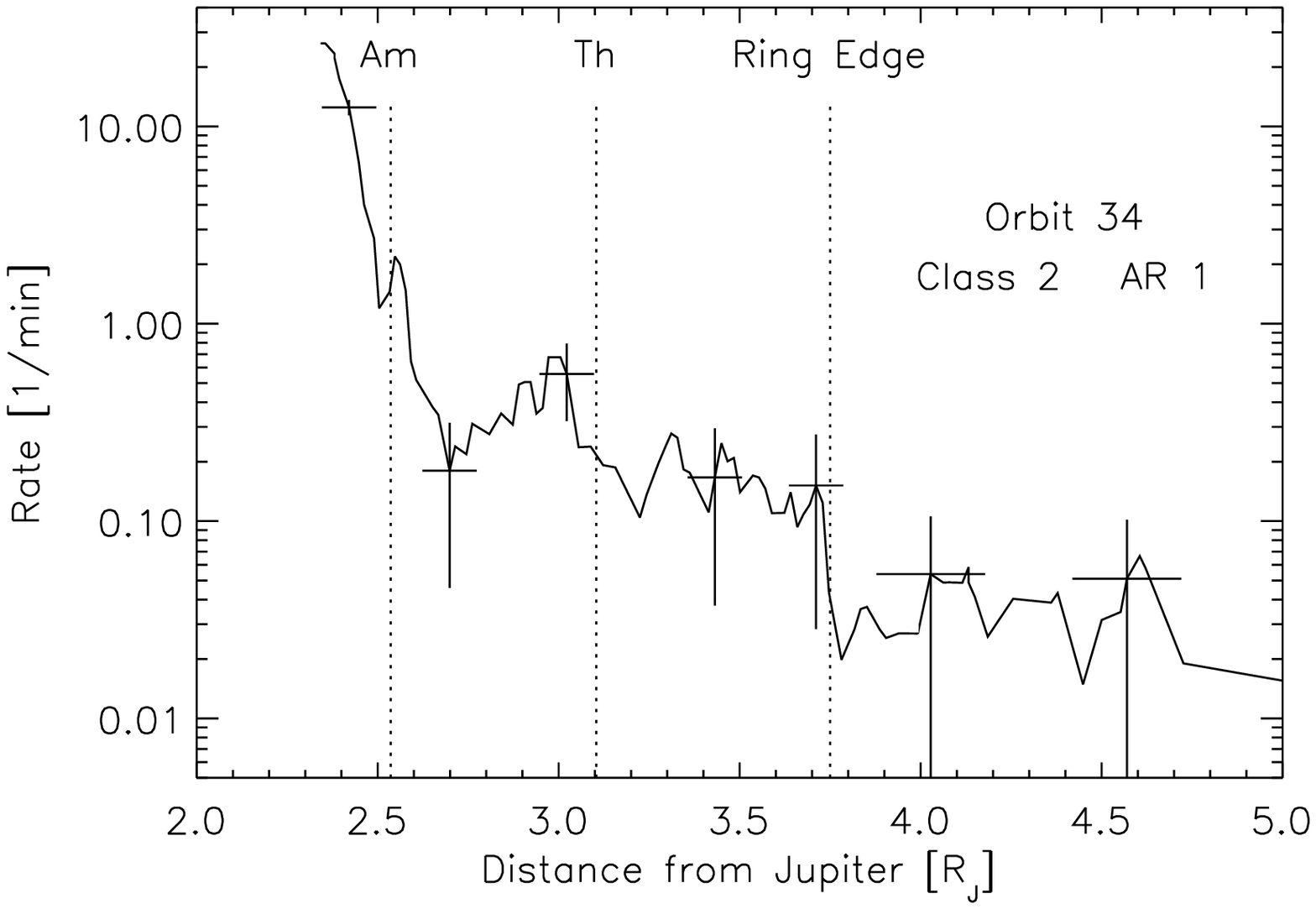}
\epsfxsize=0.5\hsize
\epsfbox{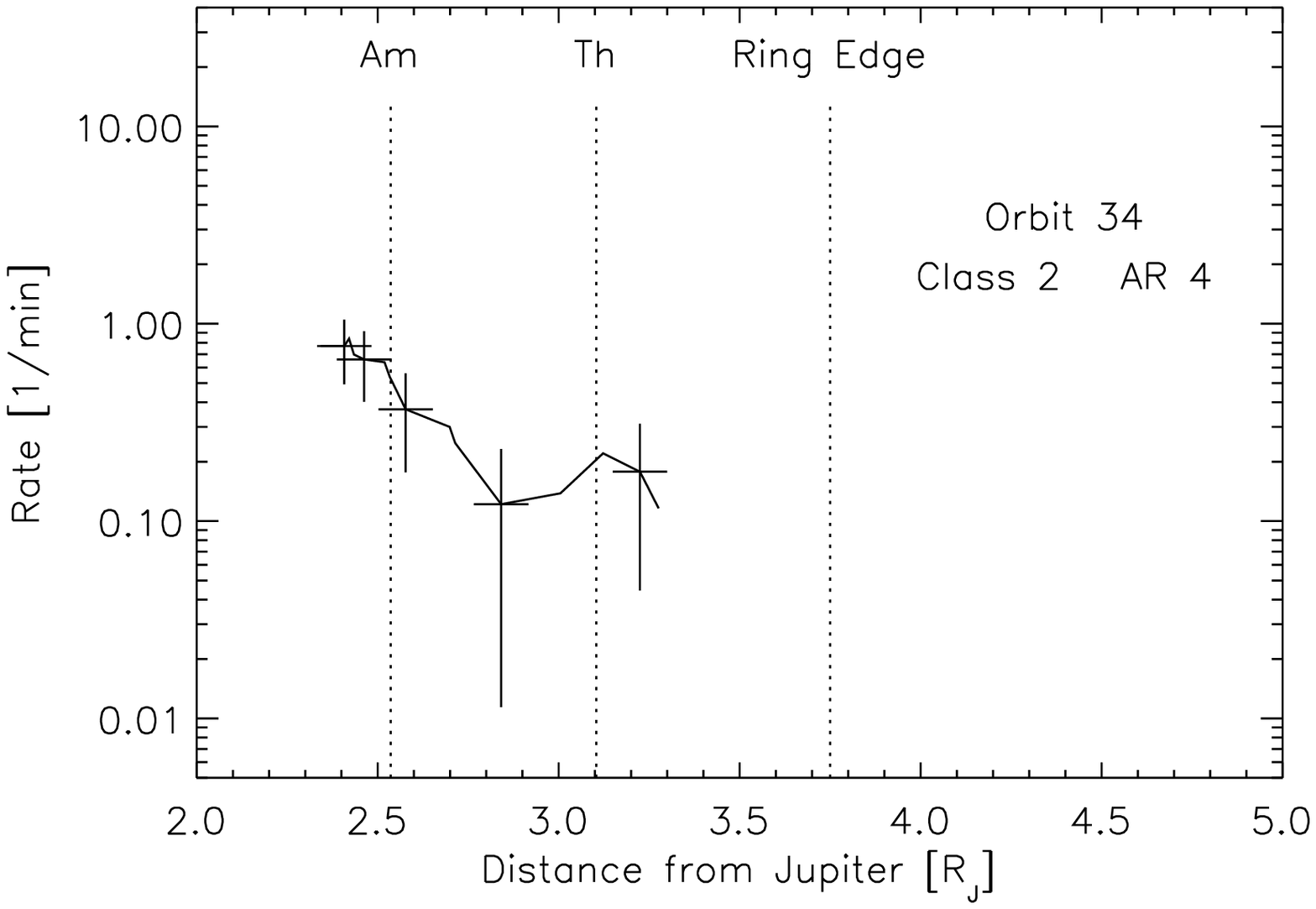}
\epsfxsize=0.5\hsize
\epsfbox{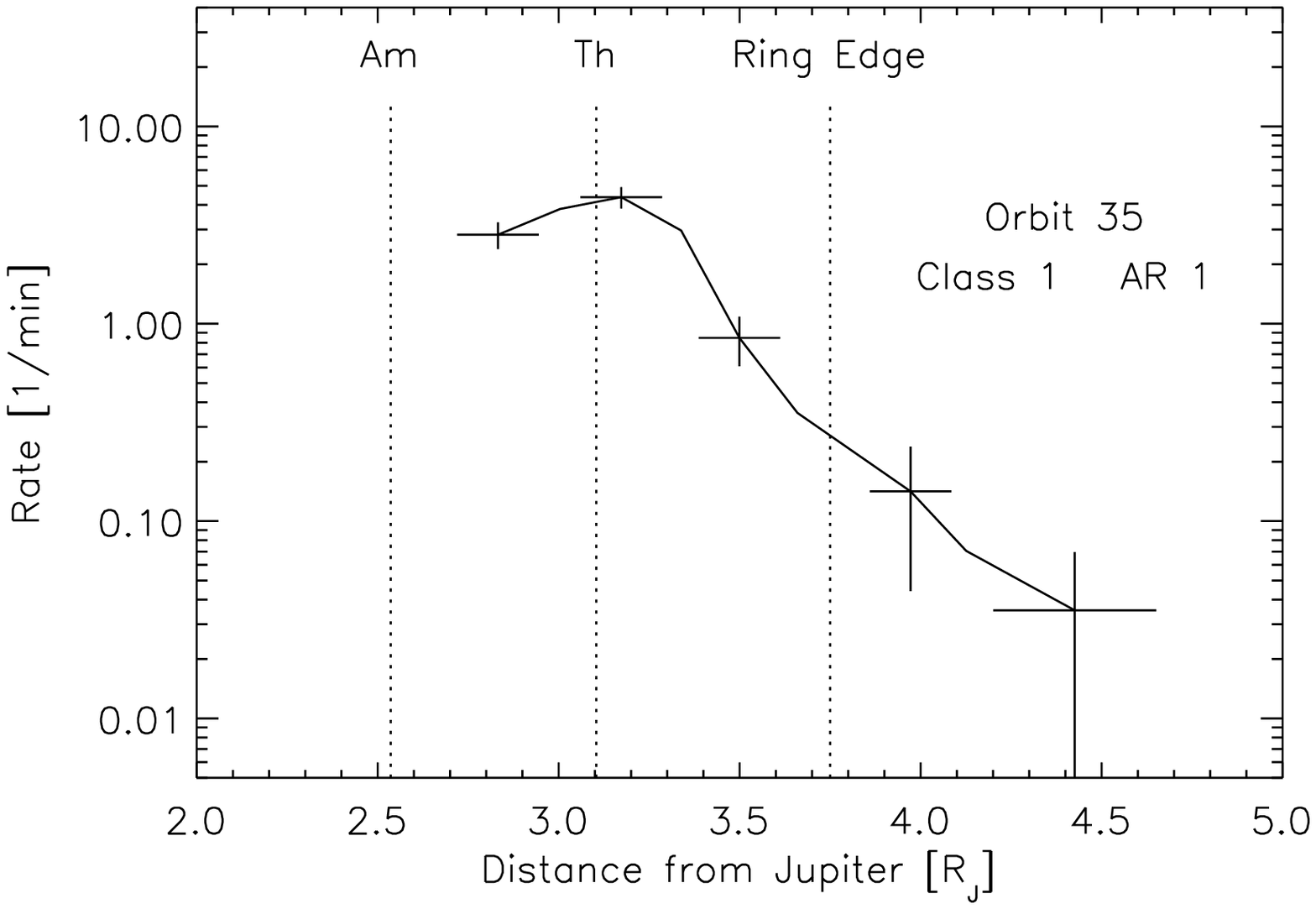}
\epsfxsize=0.5\hsize
\epsfbox{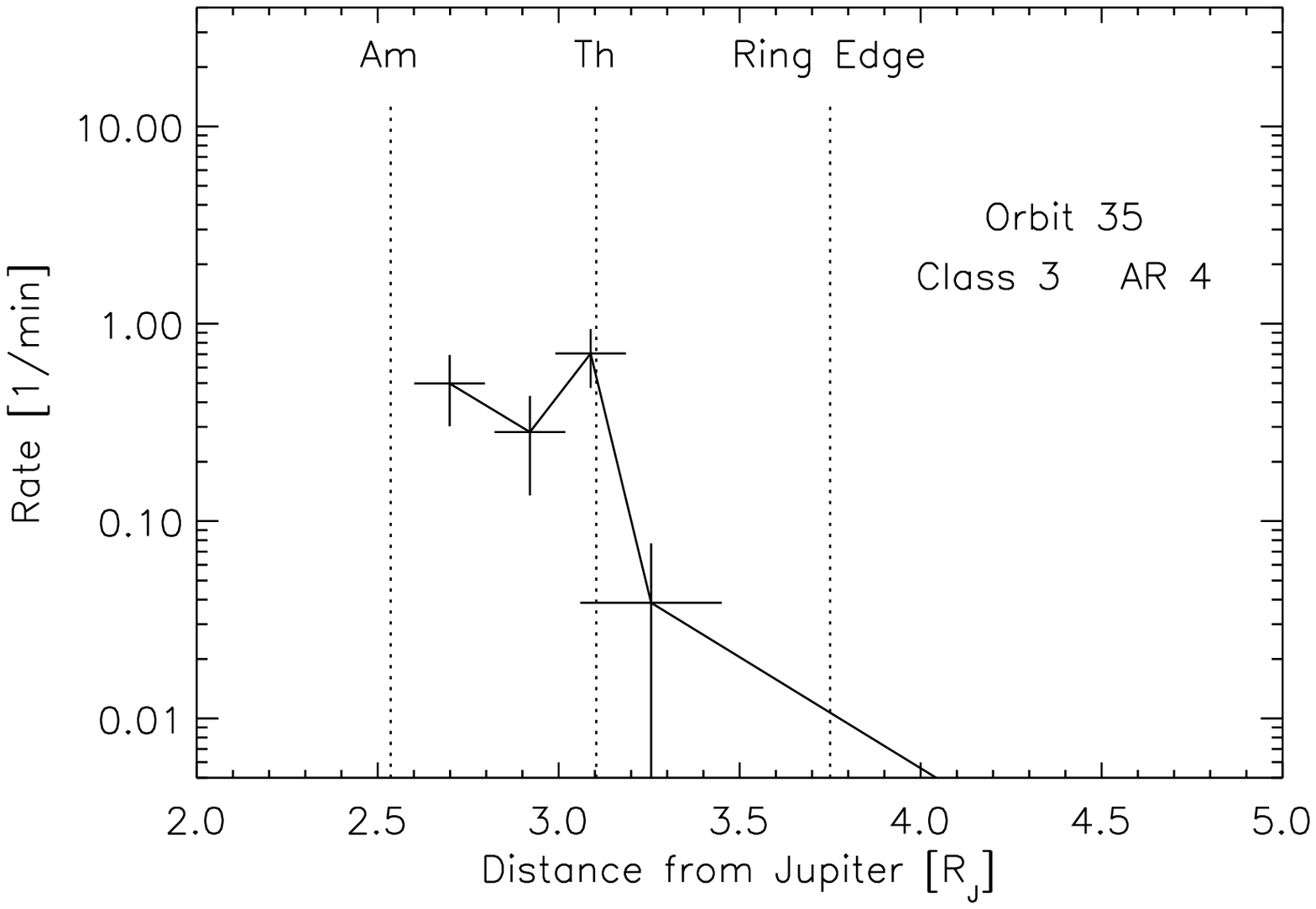}
\caption{\label{rate} Dust impact rates measured during both ring
  passages as derived from the dust instrument accumulators. The ring
  passage, class and amplitude range are given for each
  panel.  For the first ring passage (A34) data were smoothed with a
  boxcar average over 3 data points while no smoothing was applied to
  data from the second passage (J35).  Vertical dotted lines indicate
  the orbits of Amalthea ('Am'), Thebe ('Th') and the edge of the
  faint ring extension as seen on images ('Ring Edge'). Error 
  bars represent the $\sqrt{n}$ statistical fluctuation of the dust 
  impacts detected
  within a 10 to 20 minute time interval.
  }
\end{figure}

\begin{figure}
\begin{turn}{90}
\epsfxsize=0.5\hsize
\epsfbox{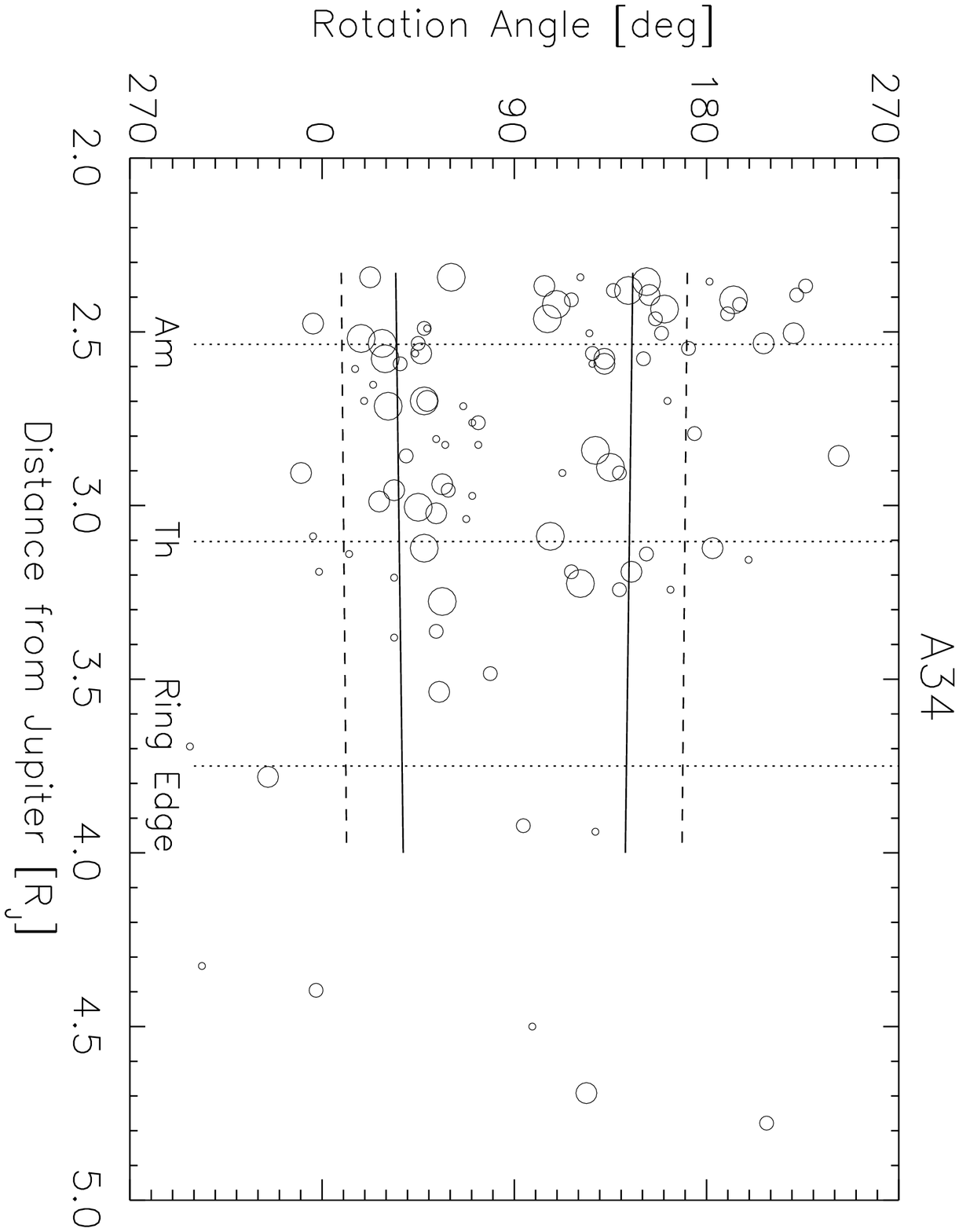}
\end{turn}
\begin{turn}{90}
\epsfxsize=0.5\hsize
\epsfbox{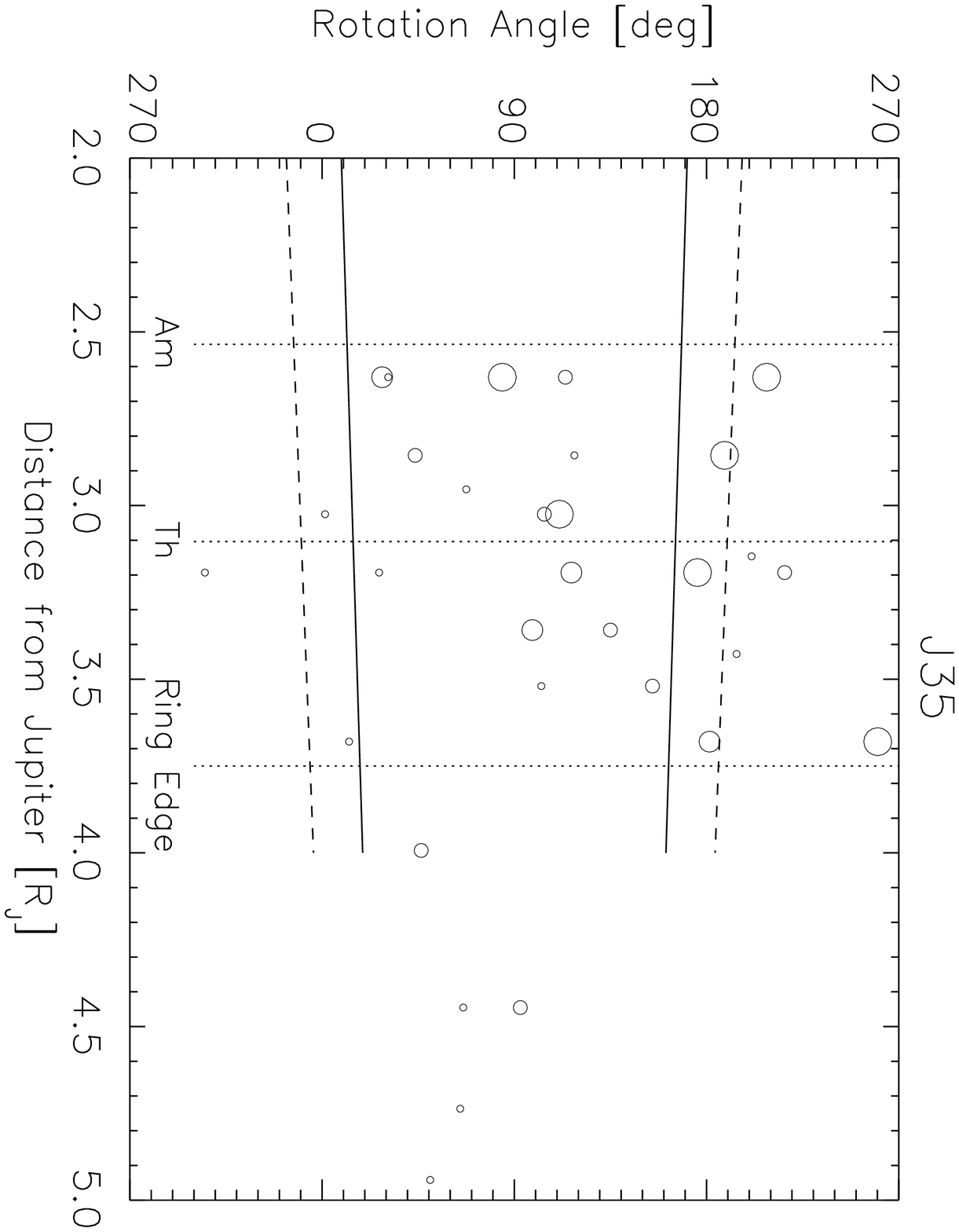}
\end{turn}
         \caption{\label{rotation_angle}
Rotation angles $\Theta$ of dust impacts measured during both gossamer ring passages.
{\em Top panel:} First ring passage on 5 November 2002.
{\em Bottom panel:} Second ring passage on 21 September 2003.
Only impacts are shown for which the complete set of measured impact 
parameters was transmitted to Earth. Solid nearly horizontal lines
indicate the expected width of the rotation angle distribution 
$\Delta \Theta$ for a sensor target $140^{\circ}$ FOV, while dashed lines
show the same for target plus sensor side wall ($180^{\circ}$ FOV).
Vertical dotted lines indicate the orbits of Io ('Io'), Thebe ('Th') and
Amalthea ('Am') and the edge of the faint ring extension as seen on
images ('Ring Edge'). 
We ignore the $1.3^{\circ}$ inclination of Jupiter's orbital plane
w.r.t. the ecliptic plane and Jupiter's obliquity of about $3^{\circ}$
and take the planet's equatorial plane to be coplanar with the
ecliptic plane for simplicity. 
}
\end{figure}

\begin{figure}
\parbox{\hsize}{
\vspace{-3cm}
\epsfxsize=0.48\hsize
\epsfbox{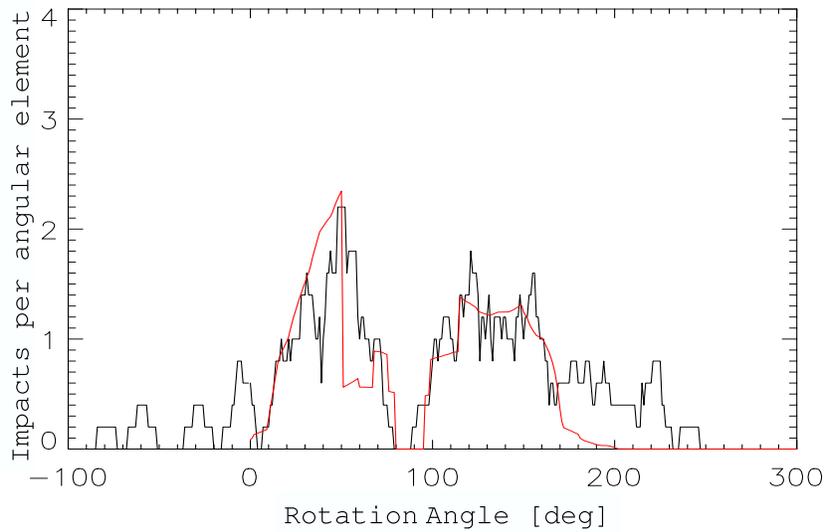}
}
\parbox{\hsize}{
\vspace{-3cm}
\epsfxsize=0.56\hsize
\epsfbox{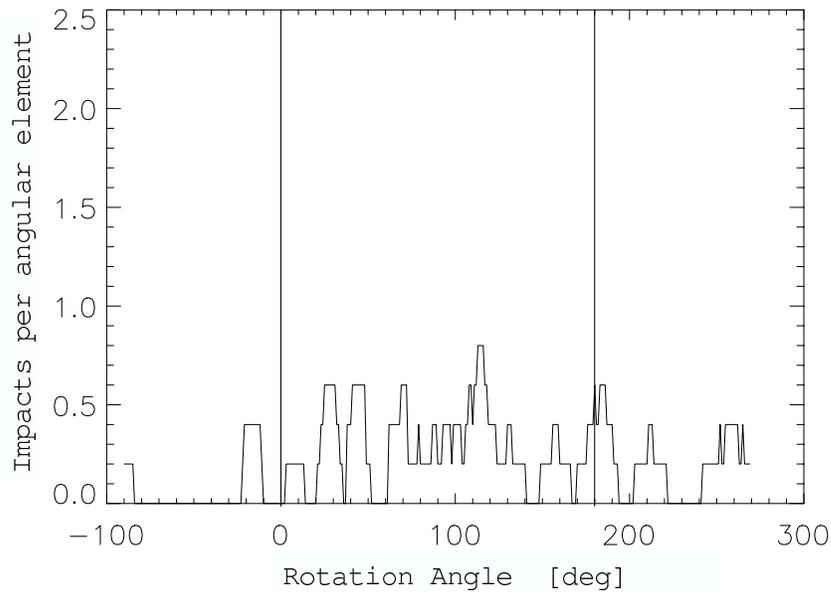}
}
\caption{\label{rot_dist_1} Distribution of rotation angles $\Theta$
  measured during both gossamer ring passages A34 (top) and J35
  (bottom). For A34 the thin solid line
  shows a modelled distribution taking into account shading by the
  magnetometer boom and the PLS and EPD instruments and inclinations
  of the particle orbits up to $20^{\circ}$
  \citep[from][]{moissl2005}. No modelling was performed for J35 
  because of the low number of detections. Here the vertical lines indicate the
  expected width from the FOV for the target and the side wall.  }
\end{figure}

The rotation angles $\Theta$ of the dust impacts measured during both
ring passages are shown in Figure~\ref{rotation_angle} and histograms
showing the number of impacts per rotation angle bin are given in
Figure~\ref{rot_dist_1}.  The rotation angle distribution measured
during the first ring passage (A34 on 5 November 2002) shows a broad
gap at $\Theta \simeq 90^{\circ}$ having a width $ \Delta \Theta
\simeq \pm 20^{\circ}$.  This is due to shadowing by the magnetometer
boom (see Fig.~\ref{ddsfov}).  No such gap in the distribution
occurred during the J35 encounter (Fig.~\ref{rotation_angle}),
consistent with the geometry of that final ring passage (Fig.~\ref{ddsfov}).

\marginpar{\fbox{Fig.~\ref{rotation_angle} and ~\ref{rot_dist_1}}}

As can be seen in Figure~\ref{rotation_angle}, the distribution of the 
rotation angles measured during the first gossamer ring passage is much 
wider than expected for a sensor target with $140^{\circ}$ FOV.
The expected width of the rotation angle distribution for particles 
on prograde circular orbits was $\Delta \Theta \simeq 100^{\circ}$ 
\citep[{\em cf.}
Figure~\ref{ddsfov}; an analysis of $\Delta \Theta $ vs.  $\psi$ -- the
angle between the impact direction and the spacecraft spin axis -- is
given by ][his Figure~2.7b]{krueger2003c}.  Hence, the
distribution of measured rotation angles $\Theta$ should cover the
range $\rm 40^{\circ} \lesssim \Theta \lesssim 140^{\circ}$.  About
half of the impacts, however, were detected with rotation angles $\rm
\Theta \gtrsim 140^{\circ}$ or $\rm \Theta \lesssim 40^{\circ}$.  If
we include the sensor side wall, the expected range widens to $\Delta
\Theta \simeq 160^{\circ}$ but is still smaller than the measured
range. A similarly extended distribution was also measured during the
second ring passage on 21 September 2003.

The rotation angle distribution shows even more structure 
than just the gap at $\Theta \simeq  90 \pm 20^{\circ}$: 
Figure~\ref{rot_dist_1} reveals an asymmetry in the sense that the
distribution with rotation angles $\Theta \geq 90^{\circ}$
is broader and shallower than the one with $\Theta \leq 90^{\circ}$. 
\citet{moissl2005} modelled the shadowing of the dust sensor FOV 
by the magnetometer boom, the PLS and EPD instruments. 
The model assumes an inclination distribution
consistent with the measured rotation angles (Figure~\ref{rotation_angle})
and a sensitive area of target plus side wall. 
A modelled curve for particles on circular jovicentric orbits 
with up to $20^{\circ}$ inclinations is shown as a grey solid line in 
Figure~\ref{rot_dist_1}. It gives an
overall good agreement with the measured distribution, in particular
considering that the spacecraft structures shading the dust
sensor are described by relatively simple approximations and that the
statistics of detected grains  is rather low. 
Deviations occur at $\Theta \sim  60 \pm 10^{\circ}$ and at 
the edge of the dust sensor FOV at  $ \Theta \grtsim 170^{\circ}$. 
In both cases the model underestimates the true number of detections. 
It has to be noted that
particularly large uncertainties occur at the edge of the FOV where the
sensitive area drops to zero.
Also,  
the modelled curve underestimates the true
width $\Delta \Theta$ of the rotation angle distribution. 
It indicates that a fraction of the detected grains
may have had orbits with even larger inclinations up to about 
$60^{\circ}$ Å and eccentricities up to
0.2 \citep{moissl2005}. In all, the particle orbits significantly
differ from the circular uninclined case implied by the ring images.


One additional potential reason for the extended rotation angle
distribution may be impacts onto the spacecraft structure close to the
dust sensor.  Impacts preferentially onto the magnetometer boom may
have generated impact plasma and secondary grain fragments which may
have hit the dust sensor, resembling true impacts at rotation angles
where direct impacts of ring particles onto the target are impossible.
Such events should have revealed their presence by peculiar impact
parameters (charge amplitudes, rise times, coincidences etc.). An
analysis of the data from both ring passages, however, did not show
evidence for such peculiarities for the majority of grains, making
this explanation unlikely \citep{moissl2005}. The extended
distribution appears, therefore, to be due to the actual distribution
of dust and implies large inclinations for many dust particles.
Inclinations of this magnitude are expected from the model of
\citet{hamilton2008}.

\subsection{Grain Masses}

\label{sec_masses}

\begin{figure}
\epsfxsize=1.0\hsize
\epsfbox{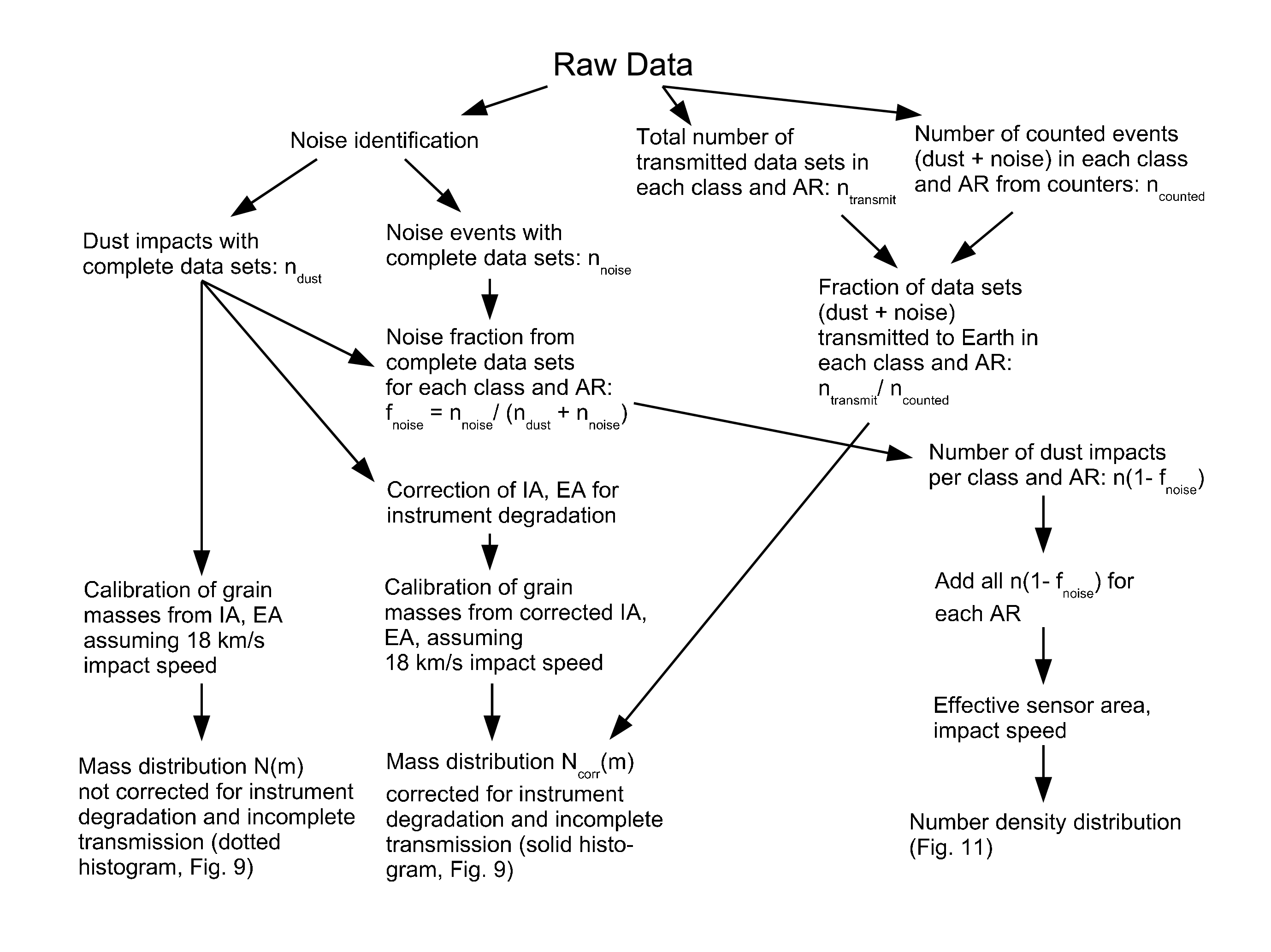}
         \caption{\label{flowchart}
Flow chart illustrating the individual processing steps required 
to derive mass and number density distributions for the 
gossamer rings.
}
\end{figure}

About 90\% of the dust impacts measured during both gossamer ring
passages showed abnormally long rise times of the impact charge signal
caused by degradation of the instrument electronics
(Section~\ref{sec_aging}).  Application of the instrument calibration
derived in the laboratory before launch would lead to unrealistically
low impact speeds and, consequently, erroneously large grain masses.
Thus, the rise time measurement cannot be used for calculating grain
impact speeds. In the gossamer rings, impact speeds are dominated by
the spacecraft's speed and, assuming that the particles move on nearly
uninclined circular orbits, the impact speed onto the detector target
on 5 November 2002 was about $\rm 18\,km\,s^{-1}$.  We use this fact
as the basis for a procedure to obtain particle mass and number
density distributions.  An overview of the individual processing steps
is given in Figure~\ref{flowchart}.

\marginpar{\fbox{Fig.~\ref{flowchart}}}

We begin by taking $\rm 18\,km\,s^{-1}$ instead of the speed derived
from the rise time measurement and calculate the particle mass with
Equation~\ref{equ_charge}, i.e. employing the linear dependence
between particle mass $m$ and impact charge $Q$. Similar mass
calibration methods were successfully applied to earlier measurements
of interstellar dust grains \citep{landgraf2000a} and to dust impacts
measured in the vicinity of the Galilean moons
\citep{krueger2000a,krueger2003b}.

An extra complication here is the amplifier degradation that arose
from the accumulated radiation damage to the dust instrument. The
damage causes the measured charge amplitude $Q$ to be too low by a
time-dependent factor that has been calculated by \citet{krueger2005a}.
For the time period of interest, we estimate the additional radiation
damage received by the spacecraft and determine a correction factor of
5 for the ion collector channel and a factor of 2 for the electron
channel, respectively.  This means that measured charges for gossamer
ring particles need to be increased by a factor of 5 and 2,
respectively, to determine the true impact charges for these channels.
Due to the linear dependence between impact charge and grain mass
(Equation~\ref{equ_charge}) this leads to an average shift in grain
mass by a factor of 3.5.

\begin{figure}
\begin{turn}{90}
\epsfxsize=0.74\hsize
\epsfbox{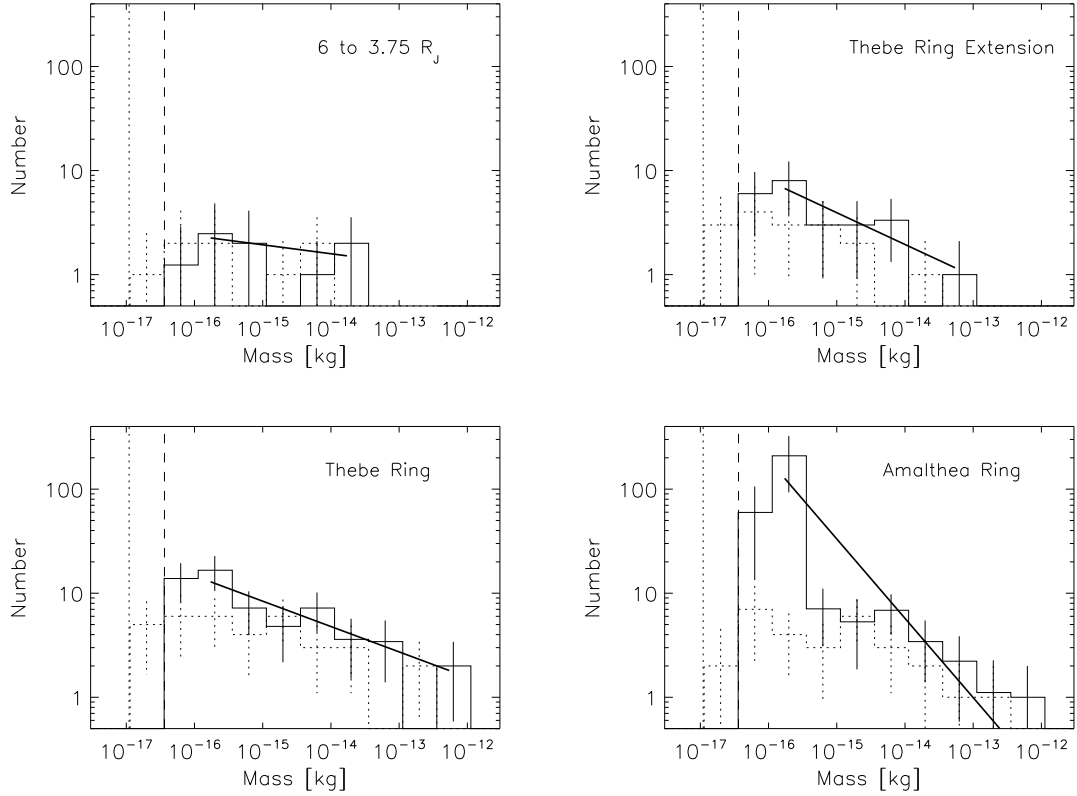}
\end{turn}
         \caption{\label{mass_hist}
Differential mass distributions per logarithmic mass interval 
for different regions of the gossamer rings
obtained from the first ring passage on 5 November 2002 
(A34; 87 particles with velocity error factor VEF $<$ 6 
\citep{gruen1995a}.
An impact speed of $\rm 18\,km\,s^{-1}$ was assumed to
calculate grain masses from the measured charge amplitudes
(Equation~\ref{equ_charge}). 
The detection threshold for the assumed impact speed is indicated by 
vertical dotted lines (without instrument aging). 
and by vertical dashed lines (with aging correction),
Dotted histograms show the distribution with neither corrections
for instrument aging nor incomplete data transmission. 
The solid histograms show the distribution corrected for 
both incomplete data transmission and instrument aging and 
thick solid lines are linear fits to these corrected
histograms. 
The slopes for the mass distributions are given in Table~\ref{massdist}.
}
\end{figure}

\marginpar{\fbox{Fig.~\ref{mass_hist}}}

In Figure~\ref{mass_hist} we show the mass distributions derived for
four different regions of the gossamer rings.  We include measurements
from: i) the region between Io's orbit and the outer edge of the Thebe
Extension (6 to $\rm 3.75\,R_J$), ii) the Thebe Extension (between
$\rm 3.75\,R_J$ and Thebe's orbit), iii) the Thebe ring (between
Thebe's and Amalthea's orbit), and iv) the Amalthea ring (inside
Amalthea's orbit). Dust in the outermost of these regions is poorly
sampled by the spacecraft and invisible from the ground. Better
statistics exist for dust amongst the Galilean satellites 
\citep{gruen1998,thiessenhusen2000,krivov2002a,krivov2002b,zeehandelaar2007}. 

To illustrate the significance of the corrections for instrument aging
and for incomplete data transmission, we show both uncorrected and
corrected histograms. The aging correction shifts the entire distribution
by a factor of 3.5 to higher masses. Coincidently, 
this corresponds to the width of half an amplitude range interval on a 
logarithmic scale so that the aging correction shifts the mass 
distribution by one histogram bin. Furthermore, to correct for
incomplete transmission, we calculated a correction factor from the
ratio between the number of counted impacts and the number of data sets
transmitted in a given time interval. We took into 
account that the leftmost two bins
correspond to AR1, the next two bins to AR2 and so on.
Note that the transmission correction is most significant
in the leftmost two bins (AR1) and nearly negligible in the other bins.

According to Figure~\ref{mass_hist} the largest detected particles
have masses $m \approx 5\times 10^{-13}\,\rm kg$.  Assuming spherical
particles with density $\rm \rho = 1000\,kg\,m^{-3}$ (representative
of water ice), the corresponding grain radius is $s \rm \simeq 5\,\mu
m$. For grain densities of 500 and $\rm 2000\,kg\,m^{-3}$ the grain
radius is 6 and $\rm 4\,\mu m$, respectively.  Similarly, the smallest
mass just exceeding the detection threshold, $m \approx 5\times
10^{-17}\,\rm kg$, corresponds to $s \rm \approx 0.2\,\mu m$.  Thus, $
0.2\,\mathrm{\mu m} \lesssim s \lesssim 5\,\mathrm{\mu m}$ is a
plausible size range from the calibration of the impact charges after
correction for electronics aging.  This shows that the size
distribution extends to particles one order of magnitude smaller than
derived from ring images.  On the other hand, the largest sizes agree
rather well with particle sizes deduced from imaging of the gossamer ring
\citep{showalter1985,showalter2008} and Jupiter's main ring
\citep{throop2004,brooks2004}. The only other
information on ring particle sizes comes from three impacts detected
at ring plane crossing by the Pioneer~10 and Pioneer~11 spacecraft
\citep{humes1976}. The Pioneer
10 detector was sensitive to particles larger than about $\rm 6\,\mu
m$ while the Pioneer 11 detector was sensitive to particles roughly
twice as large; these early measurements first showed that there
was 10 micron dust in Jupiter's equatorial plane.

Only 20 data sets of impact events were transmitted from the second
ring passage (J35) and this low number does not allow us to derive
statistically meaningful mass distributions for the individual ring
regions. In addition, the mass calibration of these data is even more
uncertain because of the rapid degradation of the dust instrument
electronics due to accelerated radiation damage very close to Jupiter 
\citep[][their Fig.~2]{krueger2005a}.

It is evident that the mass distribution is very similar in the faint
Thebe ring extension and in the Thebe ring, while it is much steeper in the
Amalthea ring. One has to keep in mind, however, that this steeper
slope is dominated by the leftmost two bins of the distribution for
masses $5 \times 10^{-16}$ -- $\rm 5 \times 10^{-17}\,kg$ which
required the largest corrections for noise removal and incomplete
transmission.  Although these bins required the largest corrections
we are convinced that the strong excess in small grains is real. 

The slopes of the differential mass distributions given
by $\mathrm{d} \log N(m)/ \mathrm{d}\log m \varpropto m^{\gamma}$
(with $N(m)$ being the number of particles per logarithmic mass interval)
for the individual ring regions are listed in Table~\ref{massdist}.
While the slopes of the Thebe ring and Thebe extension are well
reproduced by power laws 
the slope for the Amalthea ring is not very well described by a power law.

\marginpar{\fbox{Tab.~\ref{massdist}}}

Note that in all histograms the
leftmost bin is lower than the next one at higher masses.  This is a
well known effect \citep[][their Fig.~6]{krueger2006a} and is most
likely due to the fact that the sensitivity threshold of the dust
instrument may not be sharp. We therefore did not include the leftmost
bin in the fitting of power law slopes to the mass distributions. 

Interestingly, the slopes tend to steepen significantly when going
from the outer to the inner ring regions. This is due to the weakening
of electromagnetic forces in the vicinity of synchronous orbit ($\rm
2.25\,R_J$) - small particles that are expelled from the Thebe ring
cannot be ejected from the Amalthea ring \citep{hamilton1993a,hamilton2008}.

\begin{figure}
\epsfxsize=\hsize
\epsfbox{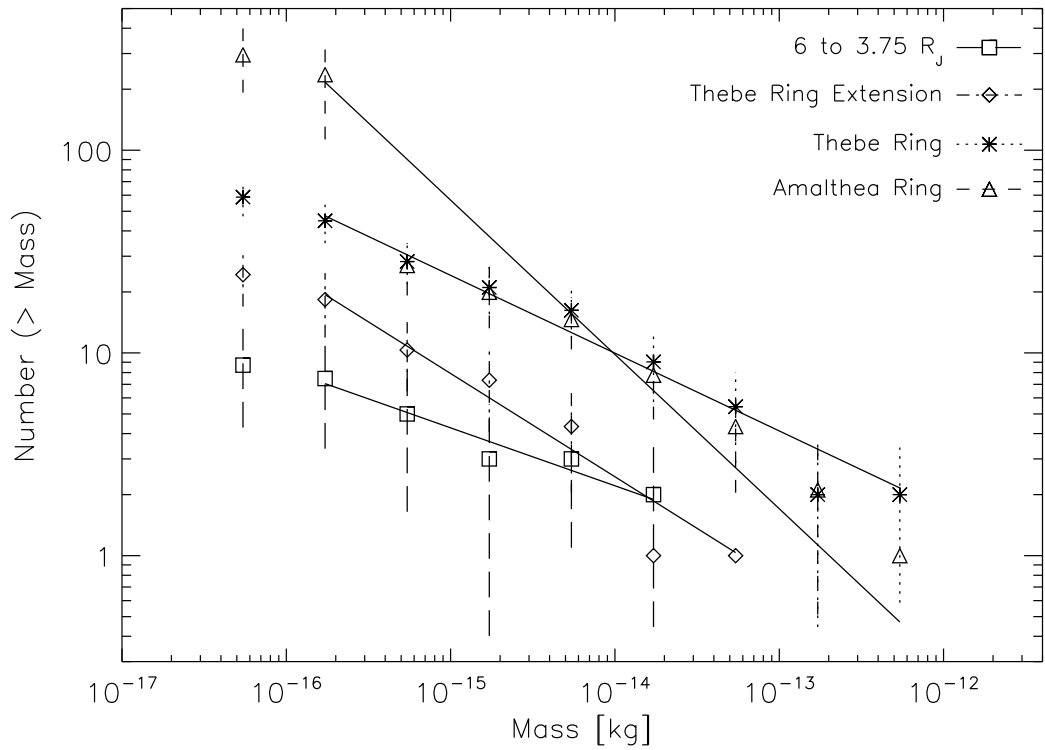}
         \caption{\label{mass_hist_cumul}
Cumulative mass distributions from Figure~\ref{mass_hist} for the
individual regions of the gossamer rings.
Straight lines are linear fits to the data.
The slopes for the cumulative mass distributions are given in Table~\ref{massdist}.
}
\end{figure}

\marginpar{\fbox{Fig.~\ref{mass_hist_cumul}}}

The cumulative mass distributions for the individual ring regions are shown
in Figure~\ref{mass_hist_cumul}. Again, the distribution for the
Amalthea ring is the steepest. The resulting power law slopes obtained
from linear fits to the data are approximately between $-0.3$ and
$-0.8$ and are tabulated in Table~\ref{massdist}.  These slopes agree
very well with the slopes measured in-situ in impact-generated dust
clouds at the Galilean moons \citep{krueger2003b}, while they are much
flatter than slopes derived for Saturn's E ring (S. Kempf, priv.
comm.). This indicates that the majority of the detected grains are
collisional ejecta from hypervelocity impacts onto the surfaces of the
moons embedded in the gossamer rings (mostly Amalthea and Thebe). 

\marginpar{\fbox{Tab.~\ref{dustpop}}}

\subsection{Dust Number Density}        \label{sec_num_dens}

Each of the impact charge amplitude ranges of the dust instrument 
corresponds to a factor of 10 in impact charge and, hence,  a factor of 10 
in mass (for constant impact speed; {\em cf.} Equation~\ref{equ_charge}). 
Therefore, a number density distribution derived from the 
accumulators directly reflects the
grain mass distribution. We use this approach to construct relative 
grain size distributions in the individual gossamer rings
without using the dust instrument calibration from the laboratory. 
The individual data processing steps are again summarised in 
Figure~\ref{flowchart}.

The dust number density $n$ is proportional to the impact rate 
$\rm d \it N /\rm d \it t$ recorded by the dust instrument,
and the relation between both quantities is given by:
\begin{equation}
n = \frac{\rm d \it N}{\rm d \it t} \cdot \frac{1}{v \cdot A_S(\psi)}
                  \label{equ_num_dens}.
\end{equation}
$A_S(\psi)$ is the sensor area as a function of the angle
$\psi$ with respect to the spacecraft spin axis, and $v$ is 
the grain impact speed. To obtain impact rates, we separated different 
ring regions into distance bins and divided the number of particles 
$\rm d \it N $ counted in a given distance bin by the time 
$\rm d \it t$ Galileo spent in this bin.

\begin{figure}
\epsfxsize=0.98\hsize
\epsfbox{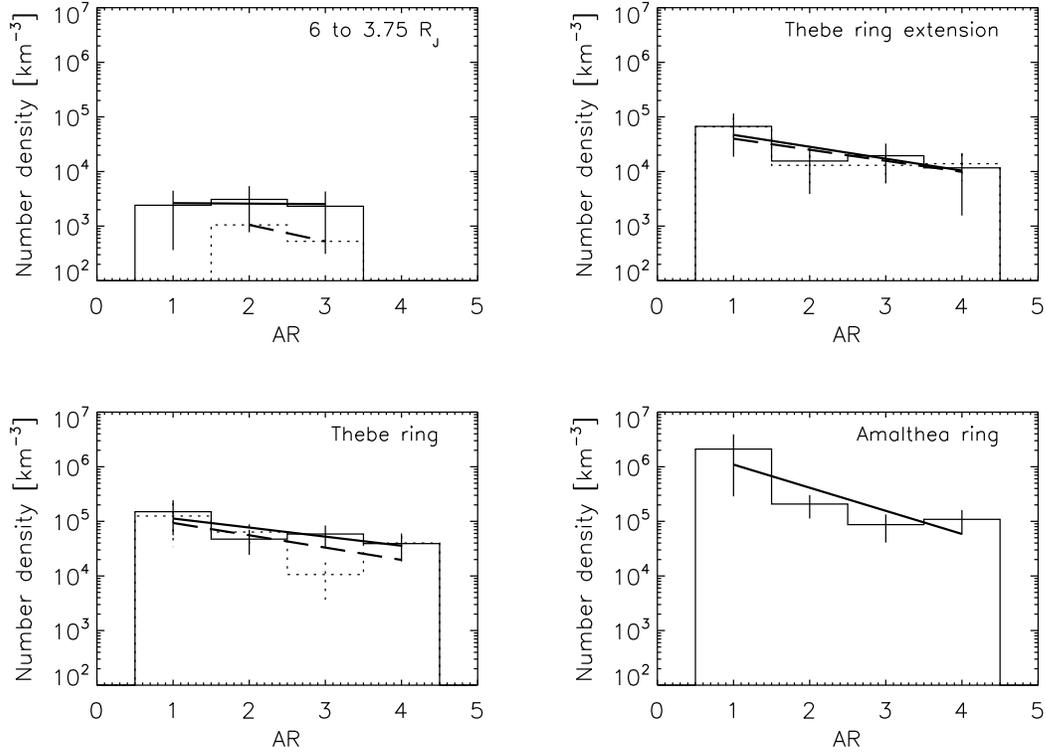}
         \caption{\label{num_dens}
Differential dust number density distributions per logarithmic mass interval
deduced from the accumulators (classes~1 to 3
taken together after noise removal). Total number densities of 
all histogram bins added in each panel as well as grain radii 
calculated with Equation~\ref{equ_radius} from these total number densities 
are given in Table~\ref{dustpop}. 
Solid lines show the data for the first gossamer ring passage (A34),
dotted lines show those for the second passage on 21 September 2003 (J35). The
assumed grain impact speeds are $\rm 18-20\,km\,s^{-1}$ and 
$\rm 26-30\,km\,s^{-1}$, respectively. 
The slopes for the number density distributions are given in Table~\ref{massdist}.
}
\end{figure}

\marginpar{\fbox{Fig.~\ref{num_dens}}}

In Figure~\ref{num_dens} we show the number densities derived from the
accumulators of the four amplitude ranges for the individual gossamer ring
regions. 
Number densities measured during both gossamer ring passages agree to
within about 50\,\%, except in the region between Io's orbit and the
outer ring edge. Here the measurements disagree by a 
factor of 3 (Figure~\ref{num_dens}).
Despite the low number of dust detections in this ring region and the
uncertainty due to the noise removal, we believe that this difference
in the number density is likely real, pointing to azimuthal variations
in the dust ring density itself. 

\citet{hamilton2008} have proposed 
that a shadow resonance governs the behavior of the gossamer rings and
their Figure 3 shows that the diffuse outer Thebe ring should be
asymmetric and offset away from the Sun. Such a structure would yield
a larger impact flux to a spacecraft approaching from the anti-Sun
hemisphere (A34, the first passage) than from the sunward hemisphere
(J35, the second passage) - see Figure~\ref{orbitplot1}. This is in qualitative
agreement with the difference in the outermost ring regions observed
here. Moreover, the Hamilton and Kr\"uger model also predicts that
larger particles should not spread very far outward from their Thebe
and Amalthea sources in agreement with the lack of AR4 grains in
Figure~\ref{num_dens} beyond the outer visible edge of the Thebe ring."

Total number densities obtained by adding the values for each
histogram bin in each panel are given in Table~\ref{dustpop}.  These
values take into account the sensor target only. If we assume that the
sensitivity of the side wall is the same as that of the target, the
number densities derived from the first ring passage are lower by
about 50\,\% while those for the second passage are reduced by only
about 10\,\%. This leads to somewhat better agreement between the two
passages. For the mass densities given in Table~\ref{dustpop} we have
assumed spherical grains with density $\rm 1000\,kg\,m^{-3}$.

In Table~\ref{dustpop} we also give number densities for dust
populations detected by Galileo beyond the orbit of Io.  Number
densities derived for the various ring regions smoothly drop with
increasing jovicentric distance, showing that Jupiter's faint ring
system fills the entire space from the gossamer rings close to Jupiter
out to the region of the Galilean moons and beyond.


\section{Discussion}

\label{sec_discussion}

\subsection{Comparison of In-Situ Data and Remote Imaging}

\label{sec_in_situ_vs_remote}

From optical imaging, ring particle size distributions can be estimated
by making assumptions about grain optical properties including the
real and imaginary components of the index of refraction and roughness
parameters. Similarly, deriving size distributions from the Galileo
dust impact data requires assumptions about instrument aging and
impact velocities. When both optical and in-situ data are available, a
new method for determining sizes is possible.

The new method has the advantage of depending only on well-measured
quantities: the ring normal optical depth, $\tau$, the ring's vertical
extension, $H$, both derived from imaging, and the number density, $n$,
measured in-situ. In particular, this calculation is independent of the
mass calibration of the dust instrument.
Relevant ring properties are given in
Table~\ref{ringproperties}.  The optical depth has the biggest error
bar whereas the ring's vertical extension is rather well known.
Furthermore, imaging shows that the rings are most tenuous
near Ju\-piter's equatorial plane and densest near their vertical
limits \citep{ockert-bell1999,depater1999}.  

\marginpar{\fbox{Tab.~\ref{ringproperties}}}

The typical ring particle radius can be expressed as
\begin{equation}
s = \sqrt{\frac{\tau}{2 \pi H n_{\rm opt}}}.      \label{equ_radius}
\end{equation}
Here, $n_{\rm opt}$ is
the number density measured in-situ of grains dominating the optical 
cross-section. But what should we use for $n_{\rm opt}$? Summing over all 
amplitude ranges yields the number densities given in Table~\ref{dustpop} 
and an effective grain radius $s \mathrm{\,\approx 2\,\mu m}$. 
In this simple analysis 
all measured particle sizes contribute to the optical cross-section.

\begin{figure}
\epsfxsize=0.98\hsize
\epsfbox{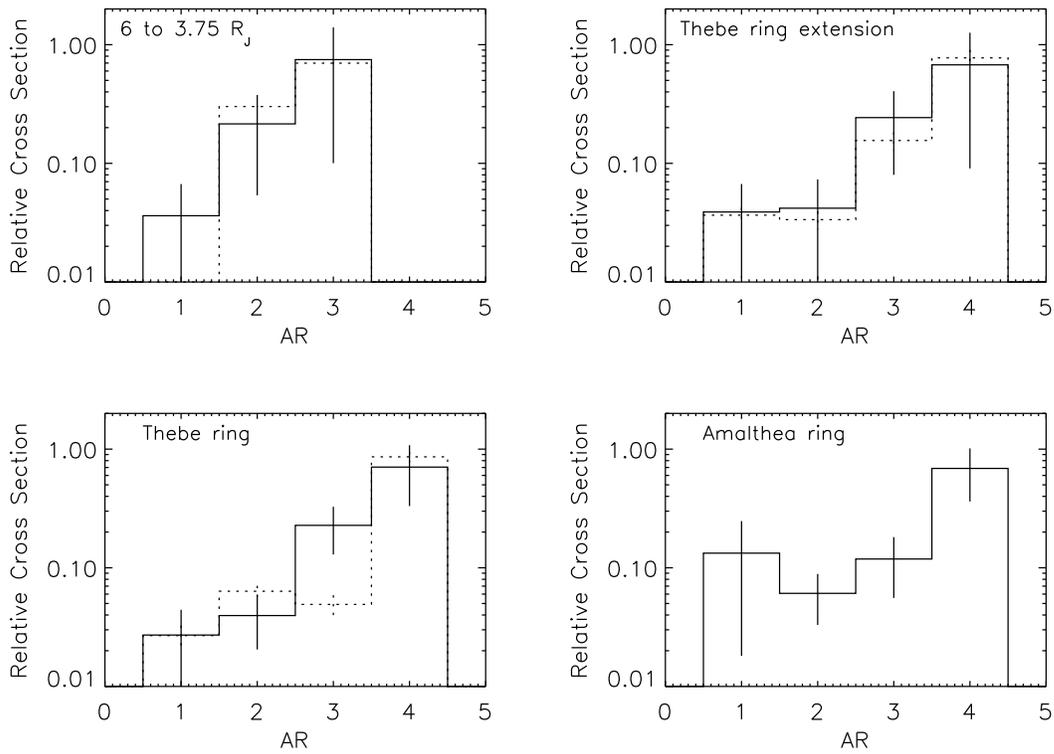}
         \caption{\label{area_dens}
Relative cross-section of dust particles deduced from the accumulators (class~1 to 3
taken together). Again, 
solid lines show the data for the first gossamer ring passage (A34),
dotted lines show those for the second passage on 21 September 2003 (J35). 
}
\end{figure}

For a more realistic calculation
we have to take into account that imaging is most sensitive
to those particles which have the largest cross-section for reflecting
light. Using the fact that amplitude ranges AR1-4
correspond to a factor of 1000 in mass (100 in area), 
Figure~\ref{area_dens} shows the relative contribution of the four
amplitude ranges to the optical cross section. 
In all ring regions the biggest
contribution to the optical depth comes from the biggest grains (AR4),
even though the smallest ones (AR1) dominate the number density.
Thus, a better choice for $n_{\rm opt}$ is to use AR4 only. 

\marginpar{\fbox{Fig.~\ref{area_dens}}}

Now taking the number densities from Figure~\ref{num_dens}
for AR4 only, the derived grain radii are $s \approx \rm 5\,\mu m$
for the Thebe ring and $ \approx \rm 10\,\mu m$ for the Amalthea ring,
respectively. Given that
the uncertainty in the optical depth is about a factor of 5 and that
of the number density is a factor of 2, we think that the grain radii
are uncertain by perhaps a factor of 3. These sizes are consistent
with the optical measurements \citep{showalter2008}, and they
agree within about a factor of 2 with the biggest sizes obtained from
the calibrated in-situ data. Given the overall uncertainties of the
dust instrument calibration and the calculation of the optical depths,
the agreement between the two methods is quite satisfactory.

%


\subsection{Grain Size Distributions}

In Sections~\ref{sec_masses} and \ref{sec_num_dens} we determined
the grain mass distributions in two different ways. Both analyses 
produced the steepest distributions in the Amalthea ring while 
further away from Jupiter the distributions are
much flatter.  However, the
slopes derived from the number density distributions 
(Section~\ref{sec_num_dens}) are 
somewhat flatter than those obtained from the mass distributions
(Section~\ref{sec_masses}, see also Table~\ref{massdist}). 
These flatter slopes are probably due to
an unsharp detection threshold of the dust instrument 
\citep{krueger2006a}, leading to an unrealistically depleted leftmost mass 
bin for the smallest particles (Figure~\ref{mass_hist}).
In order to get an estimate of the influence of this effect on the
slopes derived from the number densities, we recalculated the 
mass distributions by including all bins in the fit: 
the mass distributions became flatter, except for the Amalthea ring 
(see below), 
and they agreed very well with the slopes derived from the number 
densities. This supports our contention that the leftmost mass bin is
incomplete and should be ignored as we do in our derivation of 
column 2 of Table~\ref{massdist}. We therefore conclude that the
slopes of the mass distributions obtained from the instrument calibration
are a better measure of the true distributions in the ring than those 
derived from the number densities. 

In the Amalthea ring the fit with all bins gives a slope
of $-0.63 \pm 0.43$ for all bins which
is somewhat steeper than the slope obtained from the
number density ($-0.42 \pm 0.39$). This also
indicates that the correction for incomplete transmission 
for the Amalthea ring (which mostly affects 
the two left-most bins in the mass 
distribution) may be too strong. 

\citet{showalter2008} derived a size distribution for the
Amalthea ring which is brightest 
in the imaging. They get a power law slope of
$-2$ to $-2.5$ in the size range $4-30\,\rm \mu m$. Therefore, the in-situ
measurements and the imaging results compliment each 
other with only little overlap in the sensitive size range. 
Furthermore, a size distribution
for the main jovian ring was recently determined from Galileo 
observations by \citet{brooks2004}. They find a power law slope of
$-2.0\pm 0.3$ for particles below $\sim 15\,\mu \mathrm{m}$ and 
a transition to a power law with slope $-5.0\pm 1.5$ at larger 
sizes.

\begin{figure}
\epsfxsize=0.6\hsize
\epsfbox{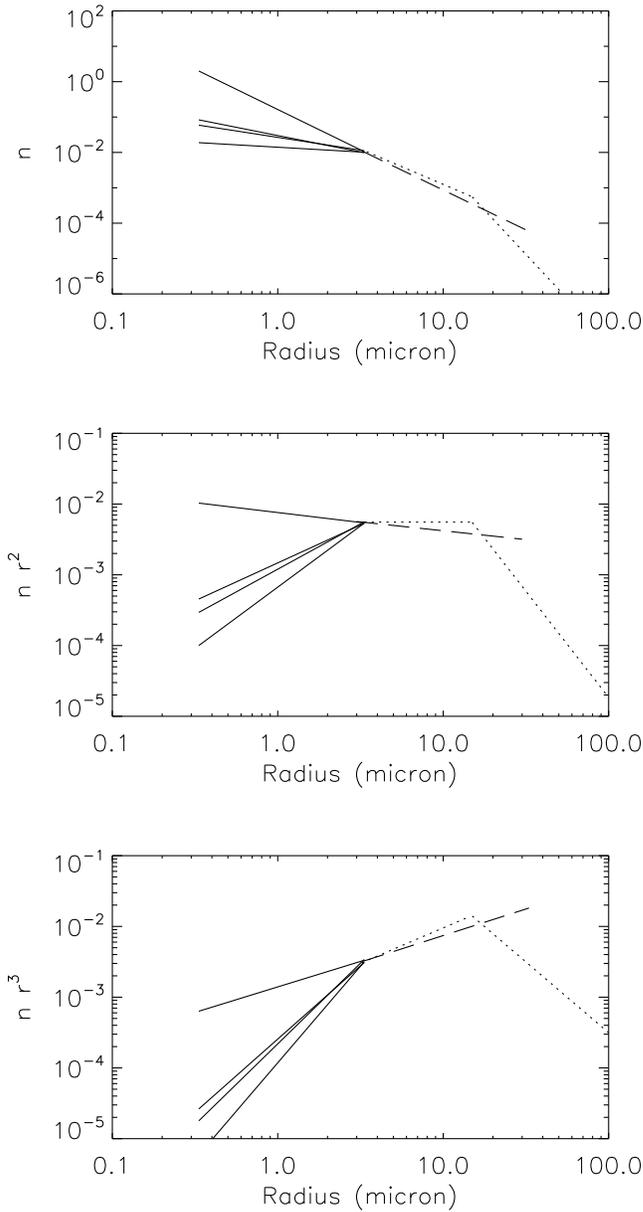}
\caption{\label{brooksplot} Relative grain size distributions per logarithmic
  size interval for the gossamer rings derived in this work (solid
  lines) compared with the ones of \citet[][dashed
  lines]{showalter2008} and \citet[][dotted lines]{brooks2004}. 
  The vertical axis is in arbitrary units and the curves are shifted so
  that they all fit together at $\mathrm{3\,\mu m}$.
  In each panel, the solid lines refer to -- from top to
  bottom -- the Amalthea ring, the Thebe ring, the Thebe extension and
  the region between the outer ring edge and Io's orbit, respectively.
  Top panel: relative 
  number density of particles in the ring; Middle panel: relative 
  cross-sectional area of the ring; Bottom panel: relative ring mass 
  density. Note that the data of \citet{brooks2004}
  are from the main jovian ring. 
 }
\end{figure}

In Figure~\ref{brooksplot} we compare these distributions with our in-situ
measurements. Note that 
the size distribution for the Amalthea ring
derived from our in-situ measurements for the small grains agrees very 
well with the one obtained from images for large grains. 
Beyond 
Amalthea's orbit the size 
distribution for submicron grains becomes flatter while little
is known about the abundance of grains bigger than 
$5\,\mu m$ in these regions.

Figure~\ref{brooksplot} is
the most complete compilation of the grain size distributions
in the jovian ring system presently available. 
It is obvious that even though the small submicron particles are the
most abundant in the rings (top panel), the largest contribution to the 
total ring mass comes from the bigger grains above $10\,\mathrm{\mu m}$ 
(bottom panel; see also Section~\ref{total_ring_mass}).

\marginpar{\fbox{Fig.~\ref{brooksplot}}}

\subsection{Total Ring Mass}

\label{total_ring_mass}

From the number density measured in-situ in the rings (Figure~\ref{num_dens})
and the known ring volume, we calculate the entire ring dust mass contained 
in the small particles ($\mathrm{0.2 - 5\,\mu m}$). Taking 
the dimensions of the Amalthea and Thebe rings
given in Table~\ref{ringproperties} and noting that the average density
near the midplane is half that of the vertical extremes,
the total mass in each of these two gossamer ring components is
about 1 to $\rm 2\times 10^6\,kg$. 
For the Thebe ring extension we find a similar value of about $\rm 10^6\,kg$ of dust, 
assuming that this ring has the same vertical extension as the Thebe ring
itself. The ring masses for the Thebe ring and Thebe ring extension
derived from Galileo's two independent ring passages agree to within
15\,\%. For the ring region between the outer edge of the Thebe ring extension
and Io's orbit we assumed the
same vertical extension as for the Thebe ring extension. Note, however, that 
there is no optical
data available for this region and dynamical simulations show that the ring 
is likely further extended. Therefore, the derived
ring mass of $\rm \approx 5\times 10^4\,kg$ is a lower limit. 
Furthermore, the two ring passages give results that differ by a
factor of three as discussed in Section~\ref{sec_num_dens}. This is probably 
due to the very asymmetric shape of the outermost ring \citep{hamilton2008}.
We collect these numbers in Table~\ref{dustpop}. 

In Figure~\ref{brooksplot} we compare the size distributions measured in-situ
(solid lines) with the ones derived from imaging (dashed and dotted lines). 
The curves are on an arbitrary scale and shifted vertically such that they fit 
together at $3\,\mu\mathrm{m}$. The bottom panel shows that the small grains 
measured in-situ represent only a minor fraction of the total ring 
mass contained in the dust: Assuming that the size distribution for 
optically visible grains in the size range 4 to $30\,\mu\mathrm{m}$ measured
by \citet{showalter2008} is valid for all gossamer rings, the total
ring mass is increased by a factor of $\sim 30$ over the values for 
small particles we list in Table~\ref{dustpop}. Similarly, if we take
the bimodal size distribution derived for the main jovian ring by 
\citet{brooks2004} in the size range
0.1 to $\rm 100\,\mu m$, the gossamer ring mass 
increases by a factor of $\sim 25$.

\subsection{Grain Dynamics}

\label{sec_dynamics}

The interesting properties of the gossamer rings can be most easily
explained with the shadow resonance model of \citet{hamilton2008}.
The shadow resonance is an electromagnetic effect that occurs
when a dust grain enters Jupiter's
shadow, photoelectric charging by solar radiation switches off, and
the grain's electric potential decreases. This leads to an 
oscillating particle charge due to the switch on
and off of photoelectric charging on the day and night side of the
planet (shadow resonance).
It changes the
electromagnetic force acting on the particle and results in coupled
oscillations of the orbital eccentricity and semimajor axis. The
oscillations cause the rings to extend significantly outward, but
only slightly inward, of their source moons while preserving their
vertical thicknesses.  This is exactly what is observed for the Thebe
ring extension. Furthermore, it leads to
longitudinally asymmetric gossamer rings, offset from the Sun for
positive grain charges and, in the absence of a dissipative drag force, 
to a lack
of material inside a certain distance from Jupiter. If most
ring material is reabsorbed by the satellites before drag forces can
draw it inward, this would create the gap interior to Thebe that is
visible in the rate plots in Figure~\ref{rate}.
\citet{showalter2008} also see evidence for a dropoff of number
density interior to Thebe's orbit. 

The existence of an at least 15000\,km wide gap in Jupiter's gossamer
ring between Thebe and Amalthea
has to be explained by the dust particle dynamics. 
Dynamical modelling by \citet{hamilton2008} shows that the shadow 
resonance, first investigated by \citet{horanyi1991}, can cause 
gaps of material interior to Thebe's orbit, lead to inclinations up to
$20^{\circ}$ for some grains, raise the fraction of small particles in 
the inner ring region, and can also explain the outward extension of the 
ring beyond the orbit of that satellite. 
It implies that electromagnetic effects have significant influence
on the dynamics of submicron- and micron-sized dust in a planetary
magnetosphere.  

An additional feature of the Galileo gossamer ring data is the likely
detection of particles on high inclination orbits.  The possibility
that spurious events, such as impacts into the detector wall or the
magnetometer boom,  masquerade as particles with high
inclinations can be most likely ruled out. Searching for a physical
explanation, the findings are consistent with grains being
driven to large inclinations by the shadow resonance as well 
\citep{hamilton2008}. The grains would
form a halo of material faint enough to be invisible to imaging, but
populated enough to be detected by direct impacts onto the Galileo
sensor. \citet{showalter2008} also see
indications for a broadening of the inclinations in the Thebe ring,
although only to a few degrees above and below the ring plane. 
Our size distribution extends to an order of magnitude smaller grains
than the smallest grains detected by the images and, thus, the expectation 
that smaller grains should be more sensitive to the shadow resonance
and thus on higher inclination orbits would be consistent with our
Galileo in-situ data. One would expect, however, that the smaller 
grains show a wider distribution in rotation angles than the bigger
ones which is in fact visible in the in-situ data: The impacts 
measured in AR4 during the A34 passage can mostly be explained 
with uninclined circular orbits while AR1 and AR3 need
orbit inclinations up to $20^{\circ}$. This is not confirmed by
the J35 data which may be due to the low number of detections.

The shadow resonance
turned out to be crucial for the structure and dust transport in 
Jupiter's tenuous dusty ring. 
Because dust from a single source can be dispersed
widely both inside and outside the source, the same mechanism may be
responsible for the wide outward extension of Saturn's E ring recently
detected with the Cassini dust instrument out to at least $\rm
18\,R_S$ (R. Srama, priv. comm.; Saturn radius $\rm R_S =
60,280\,km$) or its unexpectedly large vertical extension recently
seen on Cassini images \citep{ingersoll2007}.  
In that ring, Saturn's moon Enceladus turned out to be
the major source of ring material \citep{spahn2006a,spahn2006b}.

\section{Conclusions}

The Galileo in-situ dust detector made the first successful measurements of 
submicron and
micron-sized dust impacts in Jupiter's gossamer rings during 
two ring passages of the spacecraft in 2002 and 2003. Dust impacts were 
measured in all 
three regions of the gossamer rings which had been previously
identified on optical images. The region between Io's orbit and 
the outer limit of the faint Thebe extension,
where the ring is invisible to imaging, was also explored. The data 
from the two ring passages 
allow for the first actual comparison of in-situ dust measurements 
with the properties inferred from inverting optical images.

The measured impact rate profile 
shows a drop immediately interior to Thebe's orbit and the
grain impact directions extend over a significantly wider range than
expected for grains moving about Jupiter on uninclined circular
orbits. In fact, inclinations up to $20^{\circ}$ nicely
explain the measured impact directions for most grains. We 
investigated the idea
that spurious events, such as impacts onto the
magnetometer boom, masquerade as particles with high inclinations,
and are convinced that such explanations can be ruled out. 

The wide range in impact directions can be explained by a 
shadow resonance caused by varying particle charge on the day and 
night side of Jupiter, driving particles onto high inclination 
orbits. They form a halo of material faint enough to be invisible
to imaging, but populated enough to be detectable with the
Galileo sensor. The faint gossamer ring extension previously 
imaged to about $\rm 3.75\,R_J$ was detected out to at least 
$\rm 5\,R_J$, indicating that ejecta from Thebe spread much further
and particle orbits get higher eccentricities than previously known. 
Both the gap in the ring and the faint ring extension indicate
that the grain dynamics is strongly influenced by electromagnetic
forces.

The measured grain sizes range from about 0.2 to $\rm 5\,\mu m$, 
increasing towards smaller particles. Our measurements 
extend the known size distribution for the gossamer rings
by a factor of ten towards smaller 
particles than previously derived from imaging. Within the 
measurement uncertainties, particles contributing
most to the optical cross-section are about $\rm 5\,\mu m$ in 
radius, in agreement with imaging results. 
The grain size distribution is 
consistent with the majority of grains being generated by
hypervelocity impacts onto the surfaces of the moons orbiting 
Jupiter in the gossamer ring region. 
While the small particles 
detected in-situ are the most abundant by number, at least an order of
magnitude more mass is contained in particles larger than $5\,\mu\mathrm{m}$ 
which -- because of their large surface areas -- also dominate ring images. 
The size distributions of grains
measured in the gossamer rings gradually flatten with increasing 
distance from Jupiter due to the more efficient
electromagnetically-induced escape of more distant grains \citep{hamilton2008}.

The Galileo in-situ measurements obtained throughout the jovian 
magnetosphere show that the dust densities in Jupiter's faint ring 
system more or less continuously drop from the region of the 
gossamer rings close to Jupiter out to the Galilean moons and beyond. 
While the inner ring regions ($\mathrm{1-3.5\,R_J}$) can be clearly seen with
imaging techniques, only in-situ spacecraft can presently detect the
much fainter dust that permeates near jovian space.

\bigskip
\bigskip

{\bf Acknowledgements}

The authors wish to thank the Galileo project at NASA/JPL for effective 
and successful mission operations. This research was supported by 
the German Bundesministerium f\"ur Bildung und Forschung through Deutsches
Zentrum f\"ur Luft- und Raumfahrt e.V. (DLR, grant 50\,QJ\,9503\,3).
Support by MPI f\"ur Kernphysik is also gratefully acknowledged.
DPH would like to acknowledge NASA grant \# NNG06GGF99G for support of
this work and the MPI f\"ur Sonnensystemforschung for a short-term
visitor's grant.

\clearpage


\begin{thebibliography}{}

\bibitem[{Altobelli} et~al., 2004]{altobelli2004a}
{Altobelli}, N., {Moissl}, R., {Kr{\"u}ger}, H., {Landgraf}, M., and
  {Gr{\"u}n}, E. (2004).
\newblock {Influence of wall impacts on the Ulysses dust detector in modelling
  the interstellar dust flux}.
\newblock {\em Planetary and Space Science}, 52:1287--1295.

\bibitem[{Anderson} et~al., 2005]{anderson2005}
{Anderson}, J.~D., {Johnson}, T.~V., {Schubert}, G., {Asmar}, S., {Jacobson},
  R.~A., {Johnston}, D., {Lau}, E.~L., {Lewis}, G., {Moore}, W.~B., {Taylor},
  A., {Thomas}, P.~C., and {Weinwurm}, G. (2005).
\newblock {Amalthea's Density Is Less Than That of Water}.
\newblock {\em Science}, 308:1291--1293.

\bibitem[{Baguhl}, 1993]{baguhl1993b}
{Baguhl}, M. (1993).
\newblock {\em {Identifikation von Staubeinschl{\"a}gen in den Daten der
  Mikrometeoriden-Detektoren an Bord der Raumsonden Ulysses und Galileo}}.
\newblock PhD thesis, Ruprecht-Karls-Universit{\"a}t Heidelberg.

\bibitem[{Baguhl} et~al., 1993]{baguhl1993a}
{Baguhl}, M., {Gr{\"u}n}, E., {Linkert}, G., {Linkert}, D., and {Siddique}, N.
  (1993).
\newblock {Identification of `small' dust impacts in the Ulysses dust detector
  data}.
\newblock {\em Planetary and Space Science}, 41:1085--1098.

\bibitem[{Brooks} et~al., 2004]{brooks2004}
{Brooks}, S.~M., {Esposito}, L.~W., {Showalter}, M.~R., and {Throop}, H.~B.
  (2004).
\newblock {The size distribution of Jupiter's main ring from Galileo imaging
  and spectroscopy}.
\newblock {\em Icarus}, 170:35--57.

\bibitem[{Burns} et~al., 2001]{burns2001}
{Burns}, J.~A., {Hamilton}, D.~P., and {Showalter}, M.~R. (2001).
\newblock {Dusty Rings and Circumplanetary Dust: Observations and Simple
  Physics}.
\newblock In {Gr{\"u}n}, E., {Gustafson}, B. A.~S., {Dermott}, S.~F., and
  {Fechtig}, H., editors, {\em Interplanetary Dust}, pages 641--725. Springer
  Verlag, Berlin Heidelberg New York.

\bibitem[{Burns} et~al., 1999]{burns1999}
{Burns}, J.~A., {Showalter}, M.~R., {Hamilton}, D.~P., {Nicholson}, P.~D., {de
  Pater}, I., {Ockert-Bell}, M.~E., and {Thomas}, P.~C. (1999).
\newblock {The formation of Jupiter's faint rings}.
\newblock {\em Science}, 284:1146--1150.

\bibitem[{de Pater} et~al., 1999]{depater1999}
{de Pater}, I., {Showalter}, M.~R., {Burns}, J.~A., {Nicholson}, P.~D., {Liu},
  M.~C., {Hamilton}, D.~P., and {Graham}, J.~R. (1999).
\newblock {Keck Infrared Observations of Jupiter's Ring System near Earth's
  1997 Ring Plane Crossing}.
\newblock {\em Icarus}, 138:214--223.

\bibitem[{Frank} et~al., 1992]{frank1992}
{Frank}, L.~A., {Ackerson}, K.~L., {Lee}, J.~A., {English}, M.~R., and
  {Pickett}, G.~L. (1992).
\newblock {The Plasma Instrumentation for the Galileo mission}.
\newblock {\em Space Science Reviews}, 60:283--307.

\bibitem[{G{\"o}ller} and {Gr{\"u}n}, 1989]{goeller1989}
{G{\"o}ller}, J.~R. and {Gr{\"u}n}, E. (1989).
\newblock {Calibration of the GALILEO/ULYSSES dust detectors with different
  projectile materials and at varying impact angles}.
\newblock {\em Planetary and Space Science}, 37:1197--1206.

\bibitem[{Gr{\"u}n} et~al., 1995]{gruen1995a}
{Gr{\"u}n}, E., {Baguhl}, M., {Hamilton}, D.~P., {Kissel}, J., {Linkert}, D.,
  {Linkert}, G., and {Riemann}, R. (1995).
\newblock {Reduction of Galileo and Ulysses dust data}.
\newblock {\em Planetary and Space Science}, 43:941--951.
\newblock Paper~I.

\bibitem[{Gr{\"u}n} et~al., 1992]{gruen1992a}
{Gr{\"u}n}, E., {Fechtig}, H., {Hanner}, M.~S., {Kissel}, J., {Lindblad},
  B.~A., {Linkert}, D., {Maas}, D., {Morfill}, G.~E., and {Zook}, H.~A. (1992).
\newblock {The Galileo dust detector}.
\newblock {\em Space Science Reviews}, 60:317--340.

\bibitem[{Gr{\"u}n} et~al., 1996]{gruen1996c}
{Gr{\"u}n}, E., {Hamilton}, D.~P., {Riemann}, R., {Dermott}, S.~F., {Fechtig},
  H., {Gustafson}, B.~A., {Hanner}, M.~S., {Heck}, A., {Hor\'anyi}, M.,
  {Kissel}, J., {Kivelson}, M., {Kr{\"u}ger}, H., {Lindblad}, B.~A., {Linkert},
  D., {Linkert}, G., {Mann}, I., {McDonnell}, J. A.~M., {Morfill}, G.~E.,
  {Polanskey}, C., {Schwehm}, G.~H., {Srama}, R., and {Zook}, H.~A. (1996).
\newblock {Dust measurements during Galileo's approach to Jupiter and Io
  encounter}.
\newblock {\em Science}, 274:399--401.

\bibitem[{Gr{\"u}n} et~al., 1998]{gruen1998}
{Gr{\"u}n}, E., {Kr{\"u}ger}, H., {Graps}, A., {Hamilton}, D.~P., {Heck}, A.,
  {Linkert}, G., {Zook}, H., {Dermott}, S.~F., {Fechtig}, H., {Gustafson}, B.,
  {Hanner}, M., {Hor\'anyi}, M., {Kissel}, J., {Lindblad}, B., {Linkert}, G.,
  {Mann}, I., {McDonnell}, J. A.~M., {Morfill}, G.~E., {Polanskey}, C.,
  {Schwehm}, G.~H., and {Srama}, R. (1998).
\newblock {Galileo observes electromagnetically coupled dust in the Jovian
  magnetosphere}.
\newblock {\em Journal of Geophysical Research}, 103:20011--20022.

\bibitem[{Hamilton} and {Burns}, 1993]{hamilton1993a}
{Hamilton}, D.~P. and {Burns}, J.~A. (1993).
\newblock {Ejection of dust from Jupiter's gossamer ring}.
\newblock {\em Nature}, 364:695--699.

\bibitem[{Hamilton} and {Kr\"uger}, 2008]{hamilton2008}
{Hamilton}, D.~P. and {Kr\"uger}, H. (2008).
\newblock {Jupiter's shadow sculpts its gossamer rings}.
\newblock {\em Nature}.
\newblock in press.

\bibitem[{Hor\'anyi} and {Burns}, 1991]{horanyi1991}
{Hor\'anyi}, M. and {Burns}, J.~A. (1991).
\newblock {Charged dust dynamics - Orbital resonance due to planetary shadows}.
\newblock {\em Journal of Geophysical Research}, 96:19283--19289.

\bibitem[{Humes}, 1976]{humes1976}
{Humes}, D.~H. (1976).
\newblock {The Jovian meteoroid environment}.
\newblock In Gehrels, T., editor, {\em Jupiter}, pages 1052--1067. Univ. of
  Arizona Press, Tucson.

\bibitem[{Ingersoll} et~al., 2007]{ingersoll2007}
{Ingersoll}, A.~P., {Dyudina}, U.~A., {Ewald}, S.~P., and {Hedman}, M. (2007).
\newblock {Possible Dust Extended From Saturnian Ring Plane or Zodiacal Light
  as Seen by Cassini ISS.}
\newblock {\em AGU Fall Meeting Abstracts}, pages B1294+.

\bibitem[{Kivelson} et~al., 1992]{kivelson1992}
{Kivelson}, M.~G., {Khurana}, K.~K., {Means}, J.~D., {Russell}, C.~T., and
  {Snare}, R.~C. (1992).
\newblock {The Galileo magnetic field investigation}.
\newblock {\em Space Science Reviews}, 60:357--383.

\bibitem[{Krivov} et~al., 2002a]{krivov2002a}
{Krivov}, A.~V., {Kr{\"u}ger}, H., {Gr{\"u}n}, E., {Thiessenhusen}, K.-U., and
  {Hamilton}, D.~P. (2002a).
\newblock {A tenuous dust ring of Jupiter formed by escaping ejecta from the
  Galilean satellites}.
\newblock {\em Journal of Geophysical Research}, 107:E1, 10.1029/2000JE001434.

\bibitem[{Krivov} et~al., 2002b]{krivov2002b}
{Krivov}, A.~V., {Wardinski}, I., {Spahn}, F., {Kr{\"u}ger}, H., and
  {Gr{\"u}n}, E. (2002b).
\newblock {Dust on the outskirts of the Jovian system}.
\newblock {\em Icarus}, 157:436--455.

\bibitem[{Kr{\"u}ger}, 2003]{krueger2003c}
{Kr{\"u}ger}, H. (2003).
\newblock {\em {Jupiter's Dust Disc, An Astrophysical Laboratory}}.
\newblock Shaker Verlag Aachen, ISBN 3-8322-2224-3.
\newblock {Habilitation Thesis Ruprecht-Karls-Universit\"at Heidelberg}.

\bibitem[{Kr{\"u}ger} et~al., 2006]{krueger2006a}
{Kr{\"u}ger}, H., {Bindschadler}, D., {Dermott}, S.~F., {Graps}, A.~L.,
  {Gr{\"u}n}, E., {Gustafson}, B.~A., {Hamilton}, D.~P., {Hanner}, M.~S.,
  {Hor\'anyi}, M., {Kissel}, J., {Lindblad}, B., {Linkert}, D., {Linkert}, G.,
  {Mann}, I., {McDonnell}, J. A.~M., {Moissl}, R., {Morfill}, G.~E.,
  {Polanskey}, C., {Schwehm}, G.~H., {Srama}, R., and {Zook}, H.~A. (2006).
\newblock {Galileo dust data from the jovian system: 1997 to 1999}.
\newblock {\em Planetary and Space Science}, 54:879--910.
\newblock Paper~VIII.

\bibitem[{Kr{\"u}ger} et~al., 2001]{krueger2001a}
{Kr{\"u}ger}, H., {Gr{\"u}n}, E., {Graps}, A.~L., {Bindschadler}, D.~L.,
  {Dermott}, S.~F., {Fechtig}, H., {Gustafson}, B.~A., {Hamilton}, D.~P.,
  {Hanner}, M.~S., {Hor\'anyi}, M., {Kissel}, J., {Lindblad}, B., {Linkert},
  D., {Linkert}, G., {Mann}, I., {McDonnell}, J. A.~M., {Morfill}, G.~E.,
  {Polanskey}, C., {Schwehm}, G.~H., {Srama}, R., and {Zook}, H.~A. (2001).
\newblock {One year of Galileo dust data from the jovian system: 1996}.
\newblock {\em Planetary and Space Science}, 49:1285--1301.
\newblock Paper~VI.

\bibitem[{Kr{\"u}ger} et~al., 1999a]{krueger1999a}
{Kr{\"u}ger}, H., {Gr{\"u}n}, E., {Hamilton}, D.~P., {Baguhl}, M., {Dermott},
  S.~F., {Fechtig}, H., {Gustafson}, B.~A., {Hanner}, M.~S., {Hor\'anyi}, M.,
  {Kissel}, J., {Lindblad}, B.~A., {Linkert}, D., {Linkert}, G., {Mann}, I.,
  {McDonnell}, J. A.~M., {Morfill}, G.~E., {Polanskey}, C., {Riemann}, R.,
  {Schwehm}, G.~H., {Srama}, R., and {Zook}, H.~A. (1999a).
\newblock {Three years of Galileo dust data: II. 1993 to 1995}.
\newblock {\em Planetary and Space Science}, 47:85--106.
\newblock Paper~IV.

\bibitem[{Kr{\"u}ger} et~al., 1999b]{krueger1999c}
{Kr{\"u}ger}, H., {Gr{\"u}n}, E., {Heck}, A., and {Lammers}, S. (1999b).
\newblock {Analysis of the sensor characteristics of the Galileo dust detector
  with collimated Jovian dust stream particles}.
\newblock {\em Planetary and Space Science}, 47:1015--1028.

\bibitem[{Kr{\"u}ger} et~al., 2005]{krueger2005a}
{Kr{\"u}ger}, H., {Gr{\"u}n}, E., {Linkert}, D., {Linkert}, G., and {Moissl},
  R. (2005).
\newblock {Galileo long-term dust monitoring in the jovian magnetosphere}.
\newblock {\em Planetary and Space Science}, 53:1109--1120.

\bibitem[{Kr{\"u}ger} et~al., 2000]{krueger2000a}
{Kr{\"u}ger}, H., {Krivov}, A.~V., and {Gr{\"u}n}, E. (2000).
\newblock {A dust cloud of Ganymede maintained by hypervelocity impacts of
  interplanetary micrometeoroids}.
\newblock {\em Planetary and Space Science}, 48:1457--1471.

\bibitem[{Kr{\"u}ger} et~al., 1999c]{krueger1999d}
{Kr{\"u}ger}, H., {Krivov}, A.~V., {Hamilton}, D.~P., and {Gr{\"u}n}, E.
  (1999c).
\newblock {Detection of an impact-generated dust cloud around Ganymede}.
\newblock {\em Nature}, 399:558--560.

\bibitem[{Kr{\"u}ger} et~al., 2003]{krueger2003b}
{Kr{\"u}ger}, H., {Krivov}, A.~V., {Srem\v{c}evi\'c}, M., and {Gr{\"u}n}, E.
  (2003).
\newblock {Galileo measurements of impact-generated dust clouds surrounding the
  Galilean satellites}.
\newblock {\em Icarus}, 164:170--187.

\bibitem[{Landgraf} et~al., 2000]{landgraf2000a}
{Landgraf}, M., {Baggeley}, W.~J., {Gr\"un}, E., {Kr\"uger}, H., and {Linkert},
  G. (2000).
\newblock {Aspects of the Mass Distribution of Interstellar Dust Grains in the
  Solar System from in situ Measurements}.
\newblock {\em Journal of Geophysical Research}, 105, no. A5:10,343--10352.

\bibitem[{Moissl}, 2005]{moissl2005}
{Moissl}, R. (2005).
\newblock {\em Galileo's Staubmessungen in Jupiters Gossamer-Ringen}.
\newblock Ruprecht-Karls-Universit{\"a}t Heidelberg.
\newblock Diplom thesis.

\bibitem[{Ockert-Bell} et~al., 1999]{ockert-bell1999}
{Ockert-Bell}, M.~E., {Burns}, J.~A., {Daubar}, I.~J., {Thomas}, P.~C.,
  {Veverka}, J., {Belton}, M. J.~S., and {Klaasen}, K.~P. (1999).
\newblock {The structure of Jupiter's ring system as revealed by the Galileo
  imaging experiment}.
\newblock {\em Icarus}, 138:188--213.

\bibitem[{Showalter} et~al., 1985]{showalter1985}
{Showalter}, M.~R., {Burns}, J.~A., {Cuzzi}, J.~N., and {Pollack}, J.~B.
  (1985).
\newblock {Discovery of Jupiter's 'gossamer' ring}.
\newblock {\em Nature}, 316:526--528.

\bibitem[{Showalter} et~al., 2008]{showalter2008}
{Showalter}, M.~R., {de Pater}, I., {Verbanac}, G., {Hamilton}, D.~P., and
  {Burns}, J.~A. (2008).
\newblock {Properties and Dynamics of Jupiter's Gossamer Rings from Galileo,
  Voyager, Hubble and Keck Images}.
\newblock {\em Icarus}.
\newblock in press.

\bibitem[{Spahn} et~al., 2006a]{spahn2006b}
{Spahn}, F., {Albers}, N., {H{\"o}rning}, M., {Kempf}, S., {Krivov}, A.~V.,
  {Makuch}, M., {Schmidt}, J., {Sei{\ss}}, M., and {Miodrag Srem{\v c}evi{\'c}}
  (2006a).
\newblock {E ring dust sources: Implications from Cassini's dust measurements}.
\newblock {\em Planetary and Space Science}, 54:1024--1032.

\bibitem[{Spahn} et~al., 2006b]{spahn2006a}
{Spahn}, F., {Schmidt}, J., {Albers}, N., {H{\"o}rning}, M., {Makuch}, M.,
  {Sei{\ss}}, M., {Kempf}, S., {Srama}, R., {Dikarev}, V., {Helfert}, S.,
  {Moragas-Klostermeyer}, G., {Krivov}, A.~V., {Srem\v{c}evi\'c}, M.,
  {Tuzzolino}, A.~J., {Economou}, T., and {Gr{\"u}n}, E. (2006b).
\newblock {Cassini Dust Measurements at Enceladus and Implications for the
  Origin of the E Ring}.
\newblock {\em Science}, 311:1416--1418.

\bibitem[{Thiessenhusen} et~al., 2000]{thiessenhusen2000}
{Thiessenhusen}, K.-U., {Kr{\"u}ger}, H., {Spahn}, F., and {Gr{\"u}n}, E.
  (2000).
\newblock {Dust grains around Jupiter -- The observations of the Galileo Dust
  Detector}.
\newblock {\em Icarus}, 144:89--98.

\bibitem[{Throop} et~al., 2004]{throop2004}
{Throop}, H.~B., {Porco}, C.~C., {West}, R.~A., {Burns}, J.~A., {Showalter},
  M.~R., and {Nicholson}, P.~D. (2004).
\newblock {The jovian rings: new results derived from Cassini, Galileo, Voyager
  and Earth-based observations}.
\newblock {\em Icarus}, 172:59--77.

\bibitem[{Williams} et~al., 1992]{williams1992}
{Williams}, D.~J., {McEntire}, R.~W., {Jaskulek}, S., and {Wilken}, B. (1992).
\newblock {The Galileo Energetic Particles Detector}.
\newblock {\em Space Science Reviews}, 60:385--412.

\bibitem[{Willis} et~al., 2005]{willis2005}
{Willis}, M.~J., {Burchell}, M., {Ahrens}, T.~J., {Kr\"uger}, H., and {Gr\"un},
  E. (2005).
\newblock {Decreased values of cosmic dust number density estimates in the
  solar system}.
\newblock {\em Icarus}, 176:440--452.

\bibitem[{Willis} et~al., 2004]{willis2004}
{Willis}, M.~J., {Burchell}, M., {Cole}, M., and {McDonnell}, J. A.~M. (2004).
\newblock {Influence of impact ionization detection methods on determination of
  dust particle flux in space}.
\newblock {\em Planetary and Space Science}, 52:711--725.

\bibitem[Zeehandelaar and Hamilton, 2007]{zeehandelaar2007}
Zeehandelaar, D.~B. and Hamilton, D.~P. (2007).
\newblock {A Local Source for the Pioneer~10 and 11 Circumjovian Dust
  Detections}.
\newblock In {Kr{\"u}ger, H. and Graps, A. L.}, editor, {\em Dust in Planetary
  Systems (Workshop, September 26-30 2005, Kauai, Hawaii)}, pages 103--106.
  European Space Agency, ESA Publications SP-643.

\end{thebibliography}

\vspace{4cm}

\clearpage

\renewcommand{\baselinestretch}{1.0} 

\begin{table}
\caption[Summary of Galileo dust measurements in Jupiter's gossamer rings]
{\label{tab_dustmeas}
Characteristics of Galileo gossamer ring dust measurements.
}
\small
\begin{tabular}{lcc}
                    &
                    &
                    \\[-1.7ex]
\hline
Date (Galileo orbit number) & 
5 Nov. 2002 (A34)           &
21 Sept. 2003 (J35)            \\ 
Distance range measured     &
$>2.33\,  \rm R_J$          &
$\grtsim 2.5\, \rm R_J$              \\
Measurement time within $3.75\,\rm R_J$ &
100  min                    &
60   min                       \\
Time resolution             &
1 min                       &
7 min                          \\
Spatial resolution (radial) & 
$0.015\, \rm R_J$           &
$0.2 \,  \rm R_J$              \\
Number of dust impacts counted &
$\approx 330 $              &
$\approx 260 $                 \\
Number of data sets transmitted &
90                          &
20                             \\
Dust impact speed$^{\dagger}$ &
$\rm 18 - 20\, km\,s^{-1}$  & 
$\rm 26 - 30\, km\,s^{-1}$     \\
Dust detection threshold    &
$\rm \sim 0.2\, \mu m$      &
$\rm \sim 0.2\, \mu m$         \\
\hline\\[-2.0ex]
\end{tabular}

$^{\dagger}$: Dust particles were assumed to orbit Jupiter on circular prograde 
uninclined orbits.
%

\end{table}

\begin{sidewaystable}
\caption
{\label{massdist}
Slopes $\gamma$ of the mass distributions derived in this work for the different ring regions
(1). The Galileo orbits from which these data are derived are indicated.
(2) lists the slope of the differential mass distribution as derived from the instrument calibration (Fig.~\ref{mass_hist}),
and (3) and (4) the ones obtained from the measured number densities 
(Fig.~\ref{num_dens}), respectively. (5)
lists the cumulative mass distributions obtained from the instrument calibration
(Fig.~\ref{mass_hist_cumul}). In
column 4 the slope for the region between the outer ring limit and Io's orbit
is put in parentheses because it is derived from a very low number
of detections. 
}
\small
\begin{tabular}{lcccc}   
                    &
                    \\[-1.7ex]
\hline
\hline
Population         & 
\multicolumn{3}{c}{Differential mass distribution} & 
\multicolumn{1}{c}{Cumulative mass distribution}       \\
                    &
       from         &
\multicolumn{2}{c}{from}  &
       from         \\ 
                    & 
calibration        & 
\multicolumn{2}{c}{number density}    & 
calibration        \\
                   &
A34                &
A34                &
J35                  & 
A34                 \\
\multicolumn{1}{c}{(1)} &
(2)                 &
(3)                 &
(4)                 & 
(5)                 \\[0.3ex]
\hline \\[-2.4ex]
Amalthea ring      &
$-0.76 \pm 0.51$   &
$-0.42 \pm 0.39$   & 
$-$                &
$-0.76 \pm 0.31$   \\ 
Thebe ring         &
$-0.24 \pm 0.13$   &
$-0.17 \pm 0.18$   &
$-0.23 \pm 0.42$   & 
$-0.38 \pm 0.11$   \\
Thebe ring extension   &
$-0.31 \pm 0.16$   &
$-0.22 \pm 0.22$   &
$-0.20 \pm 0.28$   & 
$-0.51 \pm 0.15$   \\
Io to ring limit    & 
$-0.09 \pm 0.18$   &
$-0.01 \pm 0.09$   &
($-0.30 \pm 0.00$)  & 
$-0.29 \pm 0.06$    \\[0.2ex]
\hline
\hline\\[-2.0ex]
\end{tabular}
\end{sidewaystable}

\clearpage

\begin{sidewaystable}
\caption
{\label{dustpop}
Physical parameters of dust populations 
(1) detected in-situ at Jupiter. 
(2) lists the radial distance range where the particles were detected, 
(3) gives typical particle radii assuming spherical particles, 
(4) and 
(5) give the derived particle number densities and mass densities in 
space, respectively, 
(6) lists the dust mass contained in small particles (0.2 to $5\,\mu\mathrm{m}$), 
and (7) gives references. 
}
\begin{tabular}{l@{\hspace{3.0mm}}c@{\hspace{3.0mm}}c@{\hspace{3.3mm}}
                 c@{\hspace{3.5mm}}c@{\hspace{3.5mm}}cl}
                    &
                    &
                    &
                    &
                    &
                    \\[-1.7ex]
\hline
\hline
Population         & 
Jovicentric        &
Particle           & 
Number             & 
Mass               &
Dust mass in       &
Reference           \\
                    & 
distance           &
radii              & 
density            & 
density            & 
small grains       &
                      \\
                    & 
($\rm R_J$)                  &
($\rm \mu m$)     & 
($\mathrm{km^{-3}}$)&
($\mathrm{kg\,m^{-3}}$) &
(kg)                &
                     \\
\multicolumn{1}{c}{(1)} &
(2)                &
(3)                &
(4)                &
(5)                & 
(6)                & 
\multicolumn{1}{c}{(7)} \\[0.3ex]
\hline \\[-2.4ex]
Amalthea ring      &
2.33 -- 2.54       &
0.2 -- 5           &
$\sim 2\times 10^{6}$ &
$\sim 4\times 10^{-18}$ & 
$\sim 10^{6}$       &
This work                   \\
Thebe ring         &
2.54 -- 3.1        &
0.2 -- 5           &
$\sim 3\times 10^{5}$ &
$\sim 10^{-18}$    & 
$\sim 2\times 10^{6}$ &
This work           \\
Thebe ring extension  &
3.1 -- 3.75        &
0.2 -- 5           &
$ \sim 10^{5}$     & 
$ \sim 4 \times 10^{-19}$ & 
$\sim 10^{6}$&
This work                \\
Io to ring limit    & 
3.75 -- 6           &
0.2 -- 2            &
$\sim 5\times 10^{3}$ &
$\sim 5\times 10^{-21}$ &
$\approx 5\times 10^{4}$ &
This work        \\
Galilean ring         &
10 -- 30             &
0.6 -- 3           &
$10^{2} - 10^{3}$&
$10^{-21} - 10^{-20}$ &
                     &
\citet{krivov2002a}   \\
Captured particles  &
10 --  20            &
$ 0.5 - 1.5 $      &
$\sim 10^{2}$     &
$\sim 10^{-21}$    &
                     &
\citet{thiessenhusen2000}     \\
Distant ring     &
$\rm \geq 50$     &
1 -- 2             &
$\sim 10^{1}$     &
$\sim 10^{-22}$    &
                     &
\citet{krivov2002b}   \\[0.2ex]
\hline
\hline\\[-2.0ex]
\end{tabular}




\end{sidewaystable}

\clearpage 

\begin{table}[tbh]
\caption{\label{ringproperties}
Properties of the gossamer rings as obtained from imaging observations 
\citep{showalter1985,ockert-bell1999,depater1999}. 
}
   \begin{tabular}{lccl}
    \hline
    \hline \\[-2.0ex]
                             & Amalthea Ring & Thebe Ring & Uncertainty \\
\hline
                             &               &                           \\[-2.0ex]
Normal Optical Depth $\tau$ & $10^{-7}$     &   $3\times 10^{-8}$& Factor of 5 \\
Ring Half-Thickness  $H$    &  1300~km      &   4400~km  & $\rm \pm 100~km$ \\
\hline
\hline
\end{tabular}
\end{table}

\clearpage

\end{document}